\pdfoutput=1

\documentclass[12pt,a4paper]{article}

\usepackage{ifthen} 
\newboolean{pdflatex}
\setboolean{pdflatex}{true} 

\newboolean{articletitles}
\setboolean{articletitles}{true} 

\newboolean{uprightparticles}
\setboolean{uprightparticles}{false} 

\newboolean{inbibliography}
\setboolean{inbibliography}{false} 

\newboolean{uselhcbcpconvention}
\setboolean{uselhcbcpconvention}{true}
\def\paperauthors{LHCb collaboration} 
\def\paperasciititle{Template for writing LHCb papers} 
\def\paperkeywords{{High Energy Physics}, {LHCb}} 
\def\papercopyright{CERN on behalf of the LHCb collaboration}
\def\paperlicence{CC-BY-4.0}
\def\paperlicenceurl{https://creativecommons.org/licenses/by/4.0/}

\usepackage[top=1in, bottom=1.25in, left=1in, right=1in]{geometry}

\usepackage{microtype}
\usepackage{lineno}  
\usepackage{xspace} 
\usepackage{overpic}

\usepackage{caption} 

\usepackage{graphicx}  
\usepackage{color}
\usepackage{colortbl}
\graphicspath{{./figs/}} 

\usepackage{amsmath} 
\usepackage{amssymb}
\usepackage{amsfonts}
\usepackage{upgreek} 

\newcommand*\patchAmsMathEnvironmentForLineno[1]{%
\expandafter\let\csname old#1\expandafter\endcsname\csname #1\endcsname
\expandafter\let\csname oldend#1\expandafter\endcsname\csname
end#1\endcsname
 \renewenvironment{#1}%
   {\linenomath\csname old#1\endcsname}%
   {\csname oldend#1\endcsname\endlinenomath}%
}
\newcommand*\patchBothAmsMathEnvironmentsForLineno[1]{%
  \patchAmsMathEnvironmentForLineno{#1}%
  \patchAmsMathEnvironmentForLineno{#1*}%
}
\AtBeginDocument{%
\patchBothAmsMathEnvironmentsForLineno{equation}%
\patchBothAmsMathEnvironmentsForLineno{align}%
\patchBothAmsMathEnvironmentsForLineno{flalign}%
\patchBothAmsMathEnvironmentsForLineno{alignat}%
\patchBothAmsMathEnvironmentsForLineno{gather}%
\patchBothAmsMathEnvironmentsForLineno{multline}%
\patchBothAmsMathEnvironmentsForLineno{eqnarray}%
}

\usepackage{hyperxmp}

\usepackage[pdftex,
            pdfauthor={\paperauthors},
            pdftitle={\paperasciititle},
            pdfkeywords={\paperkeywords},
            pdfcopyright={Copyright (C) \papercopyright},
            pdflicenseurl={\paperlicenceurl}]{hyperref}

\usepackage[all]{hypcap} 

\usepackage{xspace} 
\usepackage{upgreek}

\def\lhcb {\mbox{LHCb}\xspace}

\def\MagUp {\mbox{\em Mag\kern -0.05em Up}\xspace}

\ifthenelse{\boolean{uprightparticles}}%
{

 \def\Ppi         {\ensuremath{\uppi}\xspace}

 \def\PDelta      {\ensuremath{\Delta}\xspace}                 
 \def\PXi      {\ensuremath{\Xi}\xspace}                 
 \def\PLambda      {\ensuremath{\Lambda}\xspace}                 
 \def\PSigma      {\ensuremath{\Sigma}\xspace}                 
 \def\POmega      {\ensuremath{\Omega}\xspace}                 
 \def\PUpsilon      {\ensuremath{\Upsilon}\xspace}                 
 
 \def\PB      {\ensuremath{\mathrm{B}}\xspace}                 
                  
 \def\PD      {\ensuremath{\mathrm{D}}\xspace}

 \def\PK      {\ensuremath{\mathrm{K}}\xspace}

 \def\Pb      {\ensuremath{\mathrm{b}}\xspace}                 
 \def\Pc      {\ensuremath{\mathrm{c}}\xspace}

 \def\Pi      {\ensuremath{\mathrm{i}}\xspace}

 \def\Ps      {\ensuremath{\mathrm{s}}\xspace}

}
{

 \def\Ppi         {\ensuremath{\pi}\xspace}

 \mathchardef\PDelta="7101
 \mathchardef\PXi="7104
 \mathchardef\PLambda="7103
 \mathchardef\PSigma="7106
 \mathchardef\POmega="710A
 \mathchardef\PUpsilon="7107
                  
 \def\PB      {\ensuremath{B}\xspace}                 
                  
 \def\PD      {\ensuremath{D}\xspace}

 \def\PK      {\ensuremath{K}\xspace}

 \def\Pb      {\ensuremath{b}\xspace}                 
 \def\Pc      {\ensuremath{c}\xspace}

 \def\Pi      {\ensuremath{i}\xspace}

 \def\Ps      {\ensuremath{s}\xspace}

}

\makeatletter
\ifcase \@ptsize \relax
  \newcommand{\miniscule}{\@setfontsize\miniscule{4}{5}}
\or
  \newcommand{\miniscule}{\@setfontsize\miniscule{5}{6}}
\or
  \newcommand{\miniscule}{\@setfontsize\miniscule{5}{6}}
\fi
\makeatother

\DeclareRobustCommand{\optbar}[1]{\shortstack{{\miniscule (\rule[.5ex]{1.25em}{.18mm})}
  \\ [-.7ex] $#1$}}

\def\g      {{\ensuremath{\Pgamma}}\xspace}

\def\squark    {{\ensuremath{\Ps}}\xspace}

\def\cquark    {{\ensuremath{\Pc}}\xspace}

\def\bquark    {{\ensuremath{\Pb}}\xspace}

\def\pion   {{\ensuremath{\Ppi}}\xspace}

\def\pip    {{\ensuremath{\pion^+}}\xspace}
\def\pim    {{\ensuremath{\pion^-}}\xspace}

\def\kaon    {{\ensuremath{\PK}}\xspace}
  \def\Kbar    {{\kern 0.2em\overline{\kern -0.2em \PK}{}}\xspace}

\def\KorKbar    {\kern 0.18em\optbar{\kern -0.18em K}{}\xspace}

\def\Kp      {{\ensuremath{\kaon^+}}\xspace}
\def\Km      {{\ensuremath{\kaon^-}}\xspace}
\def\Kpm     {{\ensuremath{\kaon^\pm}}\xspace}

  \def\Dbar    {{\kern 0.2em\overline{\kern -0.2em \PD}{}}\xspace}
\def\D       {{\ensuremath{\PD}}\xspace}

\def\DorDbar    {\kern 0.18em\optbar{\kern -0.18em D}{}\xspace}

\def\Dm      {{\ensuremath{\D^-}}\xspace}

\def\Ds      {{\ensuremath{\D^+_\squark}}\xspace}
\def\Dsp     {{\ensuremath{\D^+_\squark}}\xspace}
\def\Dsm     {{\ensuremath{\D^-_\squark}}\xspace}

\def\Dsmp    {{\ensuremath{\D^{\mp}_\squark}}\xspace}

\def\Dssm    {{\ensuremath{\D^{*-}_\squark}}\xspace}

\def\B       {{\ensuremath{\PB}}\xspace}
\def\Bbar    {{\ensuremath{\kern 0.18em\overline{\kern -0.18em \PB}{}}}\xspace}

\def\BorBbar    {\kern 0.18em\optbar{\kern -0.18em B}{}\xspace}
\def\Bz      {{\ensuremath{\B^0}}\xspace}

\def\Bu      {{\ensuremath{\B^+}}\xspace}

\def\Bp      {{\ensuremath{\Bu}}\xspace}

\def\Bd      {{\ensuremath{\B^0}}\xspace}
\def\Bs      {{\ensuremath{\B^0_\squark}}\xspace}
\def\Bsb     {{\ensuremath{\Bbar{}^0_\squark}}\xspace}

  \def\Y#1S{\ensuremath{\PUpsilon{(#1S)}}\xspace}

\def\Lbar        {{\ensuremath{\kern 0.1em\overline{\kern -0.1em\PLambda}}}\xspace}
\def\LorLbar    {\kern 0.18em\optbar{\kern -0.18em \PLambda}{}\xspace}

\def\Lbbar   {{\ensuremath{\Lbar{}^0_\bquark}}\xspace}

\def\Lcbar   {{\ensuremath{\Lbar{}^-_\cquark}}\xspace}

\newcommand{\decay}[2]{\ensuremath{#1\!\to #2}\xspace}         

\def\to                 {\ensuremath{\rightarrow}\xspace}

\def\CP                {{\ensuremath{C\!P}}\xspace}

\newcommand{\dms}{{\ensuremath{\Delta m_{\squark}}}\xspace}

\newcommand{\DGs}{{\ensuremath{\Delta\Gamma_{\squark}}}\xspace}

\newcommand{\Gs}{{\ensuremath{\Gamma_{\squark}}}\xspace}

\newcommand{\phis}{{\ensuremath{\phi_{\squark}}}\xspace}
\newcommand{\betas}{{\ensuremath{\beta_{\squark}}}\xspace}

\newcommand{\mistag}{\ensuremath{\omega}\xspace}

\newcommand{\etag}{{\ensuremath{\varepsilon_{\mathrm{tag}}}}\xspace}

\newcommand{\effeff}{\ensuremath{\varepsilon_{\mathrm{eff}}}\xspace}

\def\AT#1     {\ensuremath{A_{\mathrm{T}}^{#1}}\xspace}           

\def\C#1      {\ensuremath{\mathcal{C}_{#1}}\xspace}                       
\def\Cp#1     {\ensuremath{\mathcal{C}_{#1}^{'}}\xspace}                    
\def\Ceff#1   {\ensuremath{\mathcal{C}_{#1}^{\mathrm{(eff)}}}\xspace}        
\def\Cpeff#1  {\ensuremath{\mathcal{C}_{#1}^{'\mathrm{(eff)}}}\xspace}       
\def\Ope#1    {\ensuremath{\mathcal{O}_{#1}}\xspace}                       
\def\Opep#1   {\ensuremath{\mathcal{O}_{#1}^{'}}\xspace}                    

\newcommand{\tev}{\ifthenelse{\boolean{inbibliography}}{\ensuremath{~T\kern -0.05em eV}}{\ensuremath{\mathrm{\,Te\kern -0.1em V}}}\xspace}
\newcommand{\gev}{\ensuremath{\mathrm{\,Ge\kern -0.1em V}}\xspace}
\newcommand{\mev}{\ensuremath{\mathrm{\,Me\kern -0.1em V}}\xspace}
\newcommand{\kev}{\ensuremath{\mathrm{\,ke\kern -0.1em V}}\xspace}
\newcommand{\ev}{\ensuremath{\mathrm{\,e\kern -0.1em V}}\xspace}
\newcommand{\gevc}{\ensuremath{{\mathrm{\,Ge\kern -0.1em V\!/}c}}\xspace}
\newcommand{\mevc}{\ensuremath{{\mathrm{\,Me\kern -0.1em V\!/}c}}\xspace}
\newcommand{\gevcc}{\ensuremath{{\mathrm{\,Ge\kern -0.1em V\!/}c^2}}\xspace}
\newcommand{\gevgevcccc}{\ensuremath{{\mathrm{\,Ge\kern -0.1em V^2\!/}c^4}}\xspace}
\newcommand{\mevcc}{\ensuremath{{\mathrm{\,Me\kern -0.1em V\!/}c^2}}\xspace}

\def\mum  {\ensuremath{{\,\upmu\mathrm{m}}}\xspace}

\def\fb   {\ensuremath{\mbox{\,fb}}\xspace}
\def\invfb   {\ensuremath{\mbox{\,fb}^{-1}}\xspace}

\def\ps   {\ensuremath{{\mathrm{ \,ps}}}\xspace}

\def\invps{\ensuremath{{\mathrm{ \,ps^{-1}}}}\xspace}

\def\gsim{{~\raise.15em\hbox{$>$}\kern-.85em
          \lower.35em\hbox{$\sim$}~}\xspace}
\def\lsim{{~\raise.15em\hbox{$<$}\kern-.85em
          \lower.35em\hbox{$\sim$}~}\xspace}

\newcommand{\Real}{\ensuremath{\mathcal{R}e}\xspace}
\newcommand{\Imag}{\ensuremath{\mathcal{I}m}\xspace}

\def\sqs   {\ensuremath{\protect\sqrt{s}}\xspace}

\def\ptot       {\mbox{$p$}\xspace}
\def\pt         {\mbox{$p_{\mathrm{ T}}$}\xspace}

\def\rad{\ensuremath{\mathrm{ \,rad}}\xspace}

\def\evtgen     {\mbox{\textsc{EvtGen}}\xspace}

\def\geant      {\mbox{\textsc{Geant4}}\xspace}

\def\photos     {\mbox{\textsc{Photos}}\xspace}

\def\pythia     {\mbox{\textsc{Pythia}}\xspace}

\def\tell1  {TELL1\xspace}
\def\ukl1   {UKL1\xspace}

\newcommand{\ie}{\mbox{\itshape i.e.}\xspace}

\renewcommand{\Ds}	{\texorpdfstring{\ensuremath{D_{\hspace{-0.0625em}s}}}{Ds}\xspace} 

\newcommand{\BdDK}     {\texorpdfstring{\decay{\Bz}{\Dm \Kp}}{}}
\newcommand{\BdDPi}    {\texorpdfstring{\decay{\Bz}{\Dm \pip}}{}}

\newcommand{\BdDsK}    {\texorpdfstring{\decay{\Bz}{\Dsm \Kp}}{}}
\newcommand{\BdDsPi}   {\texorpdfstring{\decay{\Bz}{\Dsm \pip}}{}}

\newcommand{\BsDsK}    {\texorpdfstring{\decay{\Bs}{\Ds^\mp K^\pm}}{}}
\newcommand{\BsDsPi}   {\texorpdfstring{\decay{\Bs}{\Dsm \pip}}{}}
\newcommand{\BsDsRho}  {\decay{\Bs}{\Dsm \rho^{+}}}

\newcommand{\BsDspKm}  {\texorpdfstring{\decay{\Bs}{\Ds^+ K^-}}{}}
\newcommand{\BsDsmKp}  {\texorpdfstring{\decay{\Bs}{\Ds^- K^+}}{}}
\newcommand{\BsbDspKm}  {\texorpdfstring{\decay{\Bsb}{\Ds^+ K^-}}{}}

\newcommand{\BsDsstPi} {\decay{\Bs}{\Dssm\pip}}

\newcommand{\DsKKPi}   {\decay{\Dsm}{\Km\Kp\pim}}
\newcommand{\DsKPiPi}  {\decay{\Dsm}{\Km\pip\pim}}

\newcommand{\DsNonRes} {\decay{\Dsm}{(KK\pi)_{\rm nonres}}}
\newcommand{\DsPhiPi}  {\decay{\Dsm}{\phi\pim}}
\newcommand{\DsPiPiPi} {\decay{\Dsm}{\pim\pip\pim}}

\newcommand{\LbLcK}    {\decay{\Lbbar}{\Lcbar \Kp}}
\newcommand{\LbLcPi}   {\decay{\Lbbar}{\Lcbar \pip}}

\newcommand{\DsK}      {\ensuremath{\Ds\hspace{-0.0625em} K}\xspace}

\renewcommand{\g}	{\ensuremath{\gamma}\xspace}
\renewcommand{\betas}	{\ensuremath{\beta_s}\xspace}
\newcommand{\deltams}	{\ensuremath{\Delta m_s}\xspace}
\renewcommand{\dms}	{\deltams}
\newcommand{\gs}	{\ensuremath{\Gamma_s}\xspace}
\newcommand{\dgs}	{\ensuremath{\Delta\Gamma_s}\xspace}

\newcommand{\f}		{\ensuremath{f}\xspace} 
\renewcommand{\fb}	{\ensuremath{\overline{\f}}\xspace} 
\newcommand{\Af}	{\ensuremath{A_{\f}}\xspace} 

\newcommand{\Abf}	{\ensuremath{\overline{A}_{\f}}\xspace}

\newcommand{\weak}	{\ensuremath{\gamma - 2\betas}\xspace}
\newcommand{\strong}	{\ensuremath{\delta}\xspace}
\newcommand{\lf}	{\ensuremath{\lambda_{\f}}\xspace}

\newcommand{\lfb}	{\ensuremath{\lambda_{\fb}}\xspace}
\newcommand{\rdsk}	{\ensuremath{r_{\DsK}}\xspace}
\newcommand{\omcl}	{\ensuremath{1-{\rm CL}}\xspace}

\ifthenelse{\boolean{uselhcbcpconvention}}%
{%
\newcommand{\Cbpar}	{\ensuremath{C_{\fb}}\xspace}
\newcommand{\Cpar}	{\ensuremath{C_{\f}}\xspace}
\newcommand{\Sbpar}	{\ensuremath{S_{\fb}}\xspace}
\newcommand{\Spar}	{\ensuremath{S_{\f}}\xspace}
\newcommand{\Dbpar}	{\ensuremath{{A_{\fb}^{\Delta\Gamma}}}\xspace}
\newcommand{\Dpar}	{\ensuremath{{A_{\f}^{\Delta\Gamma}}}\xspace}

}{%
\newcommand{\Cbpar}	{\ensuremath{C_{\fb}}\xspace}
\newcommand{\Cpar}	{\ensuremath{C_{\f}}\xspace}
\newcommand{\Sbpar}	{\ensuremath{S_{\fb}}\xspace}
\newcommand{\Spar}	{\ensuremath{S_{\f}}\xspace}
\newcommand{\Dbpar}	{\ensuremath{D_{\fb}}\xspace}
\newcommand{\Dpar}	{\ensuremath{D_{\f}}\xspace}

}%

\newcommand{\pidk}{\ensuremath{L(K/\pi)}\xspace}

\newcommand{\sfit}	{\mbox{\it sFit}\xspace}

\usepackage{cite} 
\usepackage{mciteplus}

\usepackage{longtable} 

\begin{document}

\renewcommand{\thefootnote}{\fnsymbol{footnote}}
\setcounter{footnote}{1}


\begin{titlepage}
\pagenumbering{roman}

\vspace*{-1.5cm}
\centerline{\large EUROPEAN ORGANIZATION FOR NUCLEAR RESEARCH (CERN)}
\vspace*{1.5cm}
\noindent
\begin{tabular*}{\linewidth}{lc@{\extracolsep{\fill}}r@{\extracolsep{0pt}}}
\ifthenelse{\boolean{pdflatex}}
{\vspace*{-1.5cm}\mbox{\!\!\!\includegraphics[width=.14\textwidth]{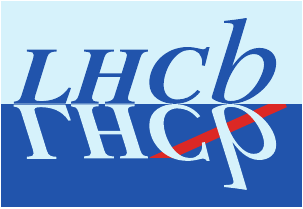}} & &}%
{\vspace*{-1.2cm}\mbox{\!\!\!\includegraphics[width=.12\textwidth]{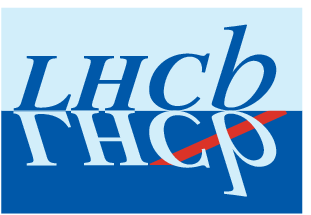}} & &}%
\\
 & & CERN-EP-2017-315 \\  
 & & LHCb-PAPER-2017-047 \\  
 & & December 20, 2017 \\ 
 & & \\
\end{tabular*}

\vspace*{4.0cm}

{\normalfont\bfseries\boldmath\huge
\begin{center}
	Measurement of \CP asymmetry in \BsDsK decays
\end{center}
}

\vspace*{2.0cm}

\begin{center}
\paperauthors\footnote{Authors are listed at the end of this paper.}
\end{center}

\vspace{\fill}

\begin{abstract}
  \noindent
   We report the  measurements of the \CP-violating parameters
in \BsDsK decays observed in $pp$ collisions, using a data set corresponding to 
	an integrated luminosity of 3.0\invfb recorded
with the \lhcb detector.
We measure  
$\Cpar  =            0.73 \pm 0.14 \pm 0.05$, 
$\Dpar  =            0.39 \pm 0.28 \pm 0.15$,
$\Dbpar =            0.31 \pm 0.28 \pm 0.15$,
$\Spar  =            -0.52 \pm 0.20 \pm 0.07$,
$\Sbpar =            -0.49 \pm 0.20 \pm 0.07$, 
where the uncertainties are statistical and 
systematic, respectively. These parameters are used together with the world-average value of the $\Bs$ mixing phase, $-2\beta_s$, to obtain a measurement of the CKM angle $\gamma$ from \BsDsK decays, yielding $\gamma =
(128\,_{-22}^{+17})^\circ$~modulo~$180^\circ$, where the uncertainty contains both statistical and systematic contributions. This corresponds to $3.8\,\sigma$ evidence for \CP violation in the interference between decay and decay after mixing.
\end{abstract}
\vspace*{2.0cm}

\begin{center}
Published in JHEP 03 (2018) 059

\end{center}

\vspace{\fill}

{\footnotesize 
\centerline{\copyright~\papercopyright, licence \href{\paperlicenceurl}{\paperlicence}.}}
\vspace*{2mm}

\end{titlepage}


\newpage
\setcounter{page}{2}
\mbox{~}

\cleardoublepage


\renewcommand{\thefootnote}{\arabic{footnote}}
\setcounter{footnote}{0}



\pagestyle{plain} 
\setcounter{page}{1}
\pagenumbering{arabic}


%

\section{Introduction}
\label{sec:intro}
A key characteristic of the Standard Model (SM) is that \CP violation originates from a single phase in the CKM quark-mixing matrix~\cite{CKM1,CKM2}. 
In the SM the CKM matrix is unitary, leading to the condition 
$V_{ud}^{\phantom{*}}V_{ub}^{*} + V_{cd}^{\phantom{*}}V_{cb}^{*} + V_{td}^{\phantom{*}}V_{tb}^{} = 0$, 
where $V_{ij}$ are the CKM matrix elements. 
This relation is represented as a triangle in the complex plane, with angles $\alpha$, $\beta$ and $\gamma$, and an area proportional 
to the amount of \CP violation in the quark sector of the SM~\cite{Jarlskog1985ht,CommentTOJarlskog1985ht,JarlskogReplyCommentTOJarlskog1985ht}. 
The angle \mbox{$\g\equiv\arg(-V^{\phantom{*}}_{ud}V_{ub}^{*}/V^{\phantom{*}}_{cd}V_{cb}^{*})$} is the least well-known angle of the CKM angles.
Its current best determination was obtained by \lhcb from a combination of measurements concerning \Bp, \Bz and \Bs decays to final states with a $D_{(s)}$ meson and one or more light mesons\cite{LHCb-PAPER-2016-032}.
Decay-time-dependent analyses of tree-level $\B^0_{(s)}\to \D^\mp_{(s)} {h^\pm}$ ($h=\pi,K$)
decays\footnote{Inclusion of charge-conjugate
modes is implied throughout except where explicitly stated.}
are sensitive to the angle $\g$ through \CP violation in the interference of mixing and decay amplitudes~\cite{Dunietz:1987bv,Aleksan:1991nh,Fleischer:2003yb,DeBruyn:2012jp}.
A comparison between the value of the CKM angle \g obtained from tree-level processes, 
with the measurements of \g and other unitary triangle parameters in loop-level processes, 
provides a powerful consistency check of the SM picture of \CP violation.

Due to the interference between mixing and decay amplitudes, the physical \CP-violating parameters in these decays
are functions of  a combination of the angle \g and the relevant mixing phase,
namely $\gamma+2\beta$ ($\beta \equiv \arg(-V^{\phantom{*}}_{cd}V_{cb}^{*}/V^{\phantom{*}}_{td}V_{tb}^{*})$)
in the \Bd and \weak
\mbox{($\betas \equiv \arg(-V^{\phantom{*}}_{ts}V_{tb}^{*}/V^{\phantom{*}}_{cs}V_{cb}^{*})$)}
in the \Bs system. Measurements of these physical
quantities can therefore be interpreted in terms of the angles \g or $\beta_{(s)}$
by using independent determinations of the other parameter as input.
Such measurements have been performed by both the BaBar~\cite{Aubert:2005yf,Aubert:2006tw}
and Belle~\cite{PhysRevD.73.092003,Bahinipati:2011yq} collaborations
using $\Bd\to D^{(*)\mp}\pi^\pm$ decays. In these decays, the 
ratios between the interfering $b \to u$ and $b \to c$ amplitudes
are small, $r_{D^{(*)}\pi} = |A(\Bd \to D^{(*)-}\pi^+)/A(\Bd \to D^{(*)+}\pi^-)| \approx 0.02$, 
which limits the
sensitivity to the CKM angle \g~\cite{HFAGSpring16}.

\begin{figure}[htb]
  \centering 
  $\,$ \hfill \includegraphics[width=.42\textwidth]{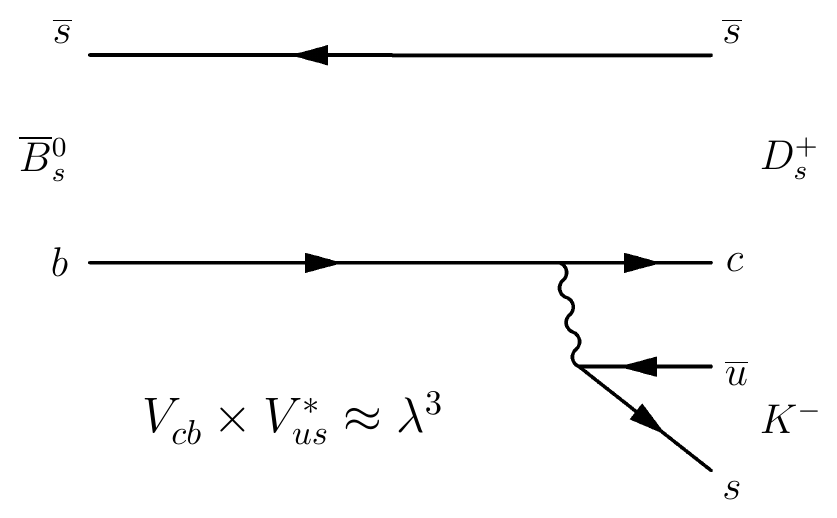} \hfill
  \includegraphics[width=.42\textwidth]{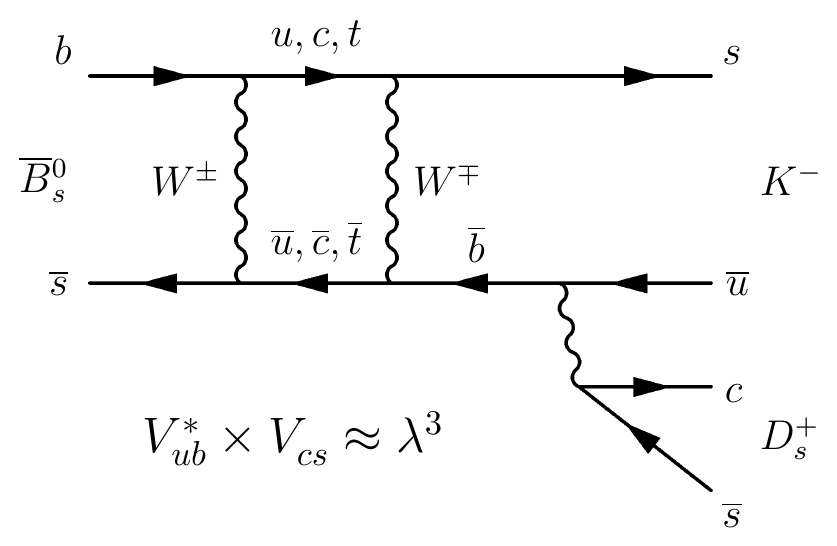} \hfill $\,$ \\
\vspace{-2mm}
  \caption{Feynman diagrams for \BsbDspKm decays (left) without and (right) with \Bs--\Bsb mixing.}
  \label{fig:feynmandiags}
\end{figure}

The leading-order Feynman diagrams contributing to the interference of decay
and mixing in \BsDsK decays are shown in Fig.~\ref{fig:feynmandiags}. 
In contrast to $\Bd\to D^{(*)\mp}\pi^\pm$ decays, here both the \BsDsmKp ($b\to cs\bar{u}$) 
and the \BsDspKm ($b\to u\bar{c}s$) decay amplitudes are of $\mathcal{O}(\lambda^3)$, 
where $\lambda \approx 0.23$~\cite{Wolfenstein:1983yz,PDG2016} is the sine of the Cabibbo angle, 
and the ratio of the amplitudes of the interfering diagrams is approximately $|V_{ub}^{*}V^{\phantom{*}}_{cs}/V^{\phantom{*}}_{cb}V_{us}^{*}| \approx 0.4$.
Moreover, the sizeable decay-width difference in the \Bs system,
\dgs~\cite{LHCb-PAPER-2013-002},
allows the determination of \weak
from the sinusoidal and hyperbolic terms of the decay-time evolution (see Eqs.~\ref{eq:decay_rates_1} and~\ref{eq:decay_rates_2}) 
up to a two-fold ambiguity.

This paper presents an updated measurement with respect to Ref.~\cite{LHCb-PAPER-2014-038} of the \CP-violating parameters and of \weak in \BsDsK decays using a data set corresponding to 
an integrated luminosity of 1.0 (2.0) \invfb of $pp$ collisions recorded with the \lhcb detector at $\sqs = 7\,(8)\tev$ in 2011 (2012).
\vspace{-2mm}

\boldmath
\subsection{Decay rate equations and \CP violation parameters}
\label{sec:equations}
\unboldmath
The time-dependent-decay rates of the initially produced flavour eigenstates
$|\Bs(t=0)\rangle$ and $|\Bsb(t=0)\rangle$ are given
by
\begin{align}
\frac{{\rm d}\Gamma_{\Bs\to\f}(t)}{{\rm d}t} &= \frac{1}{2} |\Af|^2 (1+|\lf|^2) e^{-\gs t} \left[
         \cosh\left(\frac{\dgs t}{2}\right) 
  + \Dpar\sinh\left(\frac{\dgs t}{2}\right) \right. \nonumber\\
& + \Cpar\cos\left(\dms t\right) 
  - \Spar\sin\left(\dms t\right)
\Big],
\label{eq:decay_rates_1}\\
\frac{{\rm d}\Gamma_{\Bsb\to\f}(t)}{{\rm d}t} &= \frac{1}{2} |\Af|^2 \left|\frac{p}{q}\right|^2 (1+|\lf|^2) e^{-\gs t} \left[
         \cosh\left(\frac{\dgs t}{2}\right) 
  + \Dpar\sinh\left(\frac{\dgs t}{2}\right) \right. \nonumber\\
& - \Cpar\cos\left(\dms t\right) 
  + \Spar\sin\left(\dms t\right)
\Big],
\label{eq:decay_rates_2}
\end{align}
where \mbox{$\lf \equiv (q/p)(\Abf/\Af)$} and \Af (\Abf)
is the amplitude of a
\Bs (\Bsb) decay to the final state \f,
\gs corresponds to the average \Bs decay width, 
while \dgs indicates the decay-width difference 
between the light, $|B_{L}\rangle$, and heavy, $|B_{H}\rangle$, \Bs mass eigenstates, 
defined as $\Gamma_{B_L} - \Gamma_{B_H}$ and \dms is the mixing frequency in the \Bs system defined as $m_{B_{H}} - m_{B_{L}}$. 
The complex coefficients $p$ and $q$ relate the \Bs meson mass
eigenstates,  
to the flavour eigenstates, where
\begin{equation}
\begin{aligned}
|B_L\rangle = p|\Bs\rangle+q|\Bsb\rangle \;\;\; {\text{and} }\;\;\; 
|B_H\rangle = p|\Bs\rangle-q|\Bsb\rangle\,,
\label{eq:mixing}
\end{aligned}
\end{equation}
with $|p|^2+|q|^2=1$.
Equations similar to~\ref{eq:decay_rates_1} 
and~\ref{eq:decay_rates_2} can be written 
for the decays to the \CP-conjugate 
final state $\overline{f}$ replacing
\Cpar by \Cbpar, \Spar by \Sbpar, and
\Dpar by \Dbpar. In what follows, the convention that \f (\fb) indicates $\Dsm \Kp$ ($\Dsp \Km$) final state is used.
The \CP-asymmetry parameters are given by
\begin{equation}
\begin{aligned}
\Cpar  =  \frac{ 1 - |\lf|^2 }{ 1 + |\lf|^2 } & = -\Cbpar = - \frac{1-|\lfb|^2}{1+|\lfb|^2}\,, \\
\Spar  = \frac{ 2 \Imag(\lf) }  { 1 + |\lf|^2 }        \,&,\quad
\Dpar  = \frac{ -2 \Real(\lf) }  { 1 + |\lf|^2 }        \,,  \\
\Sbpar = \frac{ 2 \Imag(\lfb) }{ 1 + |\lfb|^2 }      \,&, \quad
\Dbpar = \frac{ -2 \Real(\lfb) }{ 1 + |\lfb|^2 }      \,.
\label{eq:asymm_obs}
\end{aligned}
\end{equation}
The equality $\Cpar = -\Cbpar$ results from $|q/p| = 1$ and
$|\lf| = |1/\lfb|$, \ie assuming no \CP violation in either the mixing, in agreement with current measurements \cite{LHCb-PAPER-2016-013}, or in the decay amplitude, 
which is justified as only  a single amplitude contributes to each initial to final state transition. 
The \CP parameters are related to the
magnitude of the amplitude ratio \mbox{$\rdsk \equiv |\lambda_{\DsK}| = |A(\Bsb \to \Dsm \Kp)/A(\Bs \to \Dsm \Kp)|$},
the strong-phase difference \strong between the amplitudes $A(\Bsb \to \Dsm \Kp)$ and $A(\Bs \to \Dsm \Kp)$, and the weak-phase difference \weak by the following equations
\begin{equation}
\begin{aligned}
\Cpar  	= &\frac{1-\rdsk^2}{1+\rdsk^2}                   \,,   \\
\Dpar 	= \frac{-2 \rdsk \cos(\strong-(\weak))}{1+\rdsk^2}\,&, \quad
\Dbpar  = \frac{-2 \rdsk \cos(\strong+(\weak))}{1+\rdsk^2}\,,  \\
\Spar 	= \frac{2 \rdsk \sin(\strong-(\weak))}{1+\rdsk^2}\,&, \quad
\Sbpar	= \frac{-2 \rdsk \sin(\strong+(\weak))}{1+\rdsk^2}\,.
\label{eq:truth}
\end{aligned}
\end{equation}

\subsection{Analysis strategy}
\label{sec:strategy}

The analysis strategy
consists of a two-stage procedure.
After the event selection, an unbinned extended maximum likelihood fit, referred to as the multivariate fit, is performed to separate signal 
\BsDsK candidates from background contributions.
The multivariate fit uses the \Bs and \Dsm  invariant masses and the 
log-likelihood difference between the pion and kaon hypotheses, \pidk, for the $K^{\pm}$ candidate.
Using information from this fit, signal weights for each candidate are obtained using the \emph{sPlot} technique~\cite{Pivk:2004ty}. At the second stage, the \CP violation parameters are measured from a fit to the weighted decay-time distribution, referred to as the \sfit~\cite{2009arXiv0905.0724X} procedure, where the initial flavour of the \Bs candidate is inferred by means of several flavour-tagging algorithms optimised using data and simulation samples. 
The full procedure is validated using the flavour-specific \BsDsPi decay, yielding approximately 16 times more signal than \BsDsK decays. Precise determination of the decay-time resolution model and of the decay-time acceptance, as well as the calibration of the flavour-tagging algorithms, are obtained from \BsDsPi decays and subsequently used in the \sfit procedure to the \BsDsK candidates. 
The analysis strategy largely follows that described in Ref.~\cite{LHCb-PAPER-2014-038}. Most of the inputs are updated, in particular the candidate selection, the flavour tagging calibration and the decay-time resolution are optimised on the current data and simulation samples. A more refined estimate of the systematic uncertainties is also performed.
After a brief description of the \lhcb detector in Sec.~\ref{sec:Detector}, the event selection is reported in Sec.~\ref{sec:selection}. The relevant inputs for the multivariate fit 
and its results for \BsDsK
 and \BsDsPi decays are outlined in Secs.~\ref{sec:mdfit}. The flavour-tagging parameters and the decay-time resolution model are described in Secs.~\ref{sec:tagging} and~\ref{sec:timeres}, respectively.
The decay-time acceptance is reported in Sec.~\ref{sec:timeacc} followed by the results of the \sfit procedure applied to \BsDsK candidates in Sec.~\ref{sec:timefit}. 
The evaluation of the systematic uncertainties and the interpretation for the CKM angle \g are summarised in Secs.~\ref{sec:systematics} and~\ref{sec:interpretation}, respectively. 
Conclusions are drawn in Sec.~\ref{sec:conclusion}.

\section{Detector and software}
\label{sec:Detector}

The \lhcb detector~\cite{Alves:2008zz,LHCb-DP-2014-002} is a single-arm forward
spectrometer covering the \mbox{pseudorapidity} range $2<\eta <5$,
designed for the study of particles containing \bquark or \cquark
quarks. The detector includes a high-precision tracking system
consisting of a silicon-strip vertex detector surrounding the $pp$
interaction region~\cite{LHCb-DP-2014-001}, a large-area silicon-strip detector located
upstream of a dipole magnet with a bending power of about
$4{\mathrm{\,Tm}}$, and three stations of silicon-strip detectors and straw
drift tubes~\cite{LHCb-DP-2013-003} placed downstream of the magnet. 
The polarity of the dipole magnet is reversed periodically throughout data taking to control systematic effects. 
The tracking system provides a measurement of momentum, \ptot, of charged particles with
a relative uncertainty that varies from 0.5\% at low momentum to 1.0\% at 200\gevc.
The minimum distance of a track to a primary vertex (PV), the impact parameter (IP), 
is measured with a resolution of $(15+29/\pt)\mum$,
where \pt is the component of the momentum transverse to the beam, in\,\gevc.
Particle identification (PID) of charged hadrons is achieved using information from two ring-imaging Cherenkov detectors~\cite{LHCb-DP-2012-003}.

The online event selection is performed by a trigger~\cite{LHCb-DP-2012-004}, 
which consists of a hardware stage, based on information from the calorimeters and muon
systems, followed by a software stage, which applies a full event
reconstruction.
At the hardware trigger stage, events are required to have a muon with high \pt or a
hadron, photon or electron with high transverse energy in the calorimeters. For hadrons,
the transverse energy threshold is 3.5\gev.
The software trigger requires a two-, three- or four-track
secondary vertex with a significant displacement from any primary
$pp$ interaction vertex. At least one charged particle
must have a transverse momentum $\pt > 1.6\gevc$ and be
inconsistent with originating from any PV.
A multivariate algorithm~\cite{BBDT} is used for
the identification of secondary vertices consistent with the decay
of a \bquark hadron.

In the simulation, $pp$ collisions are generated using
\pythia~\cite{Sjostrand:2007gs, Sjostrand:2006za}  with a specific \lhcb configuration~\cite{LHCb-PROC-2010-056}.  
Decays of hadronic particles are described by \evtgen~\cite{Lange:2001uf}, in which final-state
radiation is generated using \photos~\cite{Golonka:2005pn}. 
The interaction of the generated particles with the detector, and 
its response, are implemented using the \geant
toolkit~\cite{Allison:2006ve, *Agostinelli:2002hh} as described in
Ref.~\cite{LHCb-PROC-2011-006}.

\section{Candidate selection}
\label{sec:selection}

First, \DsKKPi, \DsKPiPi, and \DsPiPiPi candidates are formed from reconstructed charged particles. These \Dsm candidates
are subsequently combined with a fourth particle, referred to as the
``companion'', to form \BsDsK or \BsDsPi candidates, depending on the PID information of the companion particle.
The decay-time resolution is improved by performing 
a kinematic fit~\cite{Hulsbergen:2005pu} in which the
\Bs candidate is assigned to a PV for which it has the smallest impact parameter $\chi^2$, defined as the difference in the $\chi^{2}$ of the vertex fit for a given PV reconstructed with and without 
the considered particle. Similarly, the \Bs invariant mass resolution is improved by constraining the \Dsm invariant mass to its world-average value.

A selection  of reconstructed candidates is made using a similar multivariate secondary-vertex algorithm as that applied at the trigger level, but with offline-quality reconstruction~\cite{BBDT}.
Combinatorial background is further suppressed by a gradient boosted decision tree (BDTG) algorithm~\cite{Breiman,Roe}, which is
trained on \BsDsPi data. 
Only the \DsKKPi final state selected with additional PID requirements is considered in order to enrich the training sample with signal candidates.
Since all channels in this analysis have similar kinematics, 
and no PID information is used as input to the BDTG,
the resulting BDTG performs equally well on the other \Dsm decay modes.
The optimal working point is chosen to maximise the significance of the \BsDsK signal.
In addition, the \Bs and \Dsm candidates
are required to have a measured mass within $[5300,5800]\mevcc$ and $[1930,2015]\mevcc$, respectively.

Finally, a combination of PID information and kinematic vetoes is used to distinguish the different 
\Dsm final states from each other (\DsKPiPi, \DsPiPiPi and \DsKKPi, the latter being subdivided into \DsPhiPi, $\Dsm \to K^{*}(892)^{0} \Km$  and \DsNonRes) and from cross-feed backgrounds such as 
\BdDK or \LbLcK decays. 
The selection structure and most criteria are identical to those used in Ref.~\cite{LHCb-PAPER-2014-038}; 
the specific values of certain PID selection requirements were updated to perform optimally with the latest event reconstruction algorithms.
Less than 1\% of the events passing the selection requirements contain more than one signal candidate.
All candidates are used in the analysis.

\boldmath
\section{Multivariate fit to \BsDsK and \BsDsPi}
\label{sec:mdfit}
\unboldmath
The signal and background probability density functions (PDFs) 
for the multivariate fit 
are obtained using a mixture of data-driven approaches and simulation.
The simulated events are corrected for differences in the transverse momentum and event occupancy distributions between simulation and data, as well
as for the kinematics-dependent efficiency of the PID selection requirements. 

The shape of the \Bs invariant mass distribution for signal candidates is modelled using the sum of two Crystal Ball functions with a common mean~\cite{Skwarnicki:1986xj}. This choice of functions provides a good description of the main peak as well as the radiative tail and reconstruction effects.
The signal PDFs are determined separately for the \BsDsK and \BsDsPi decays from simulation, taking into account different \Dsm final states.
The shapes are fixed in the nominal fit with two exceptions.
The common mean of the Crystal Ball functions is left free for both \BsDsPi and \BsDsK, compensating for differences in the mass reconstruction between simulation and data.
A scale factor accounting for data-simulation differences in the signal width is left free in the \BsDsPi fit and
is subsequently fixed to its measured value in the fit to the \BsDsK sample.
\begin{figure}[htb!]
  \centering
  \includegraphics[width=.485\textwidth]{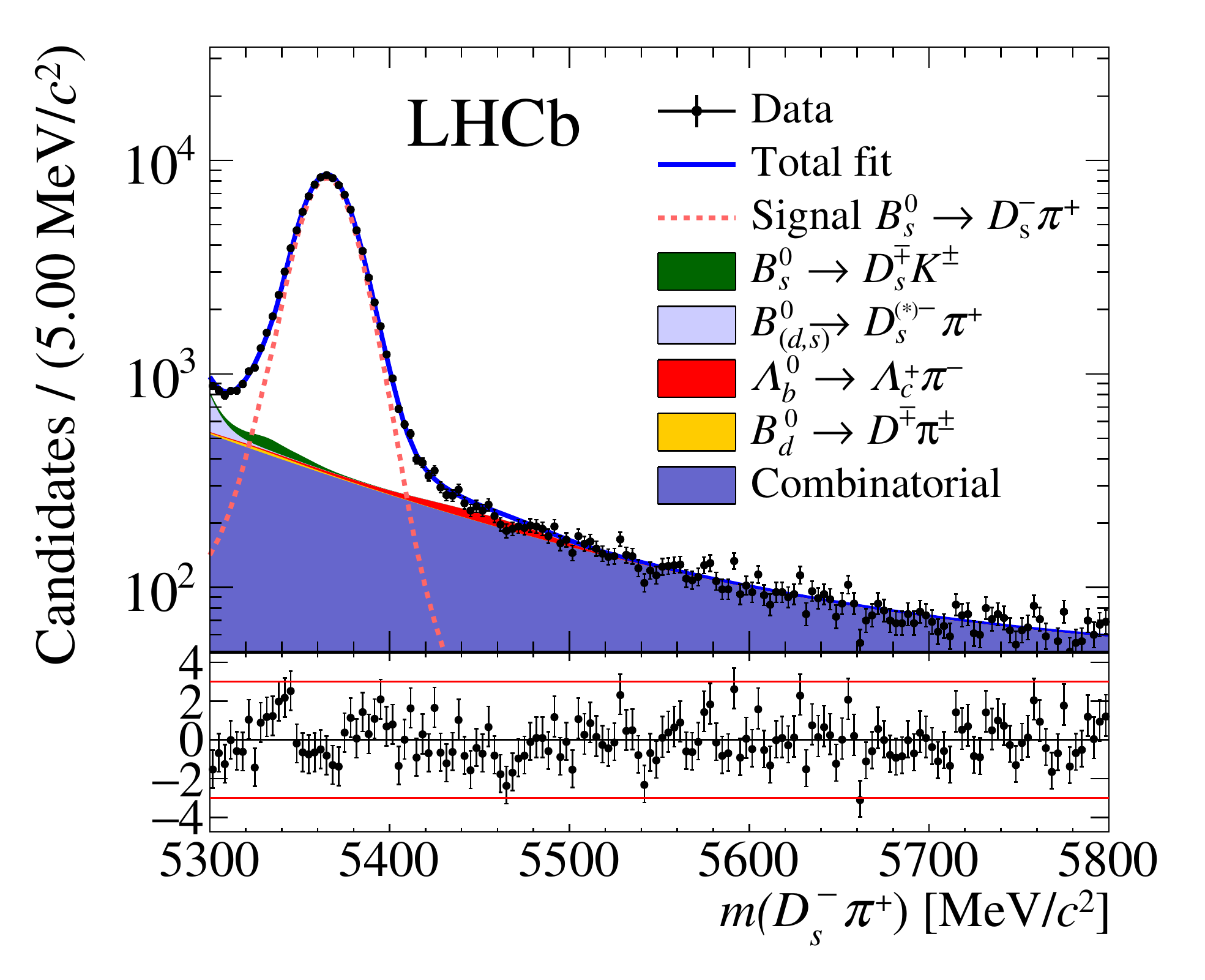}
  \includegraphics[width=.485\textwidth]{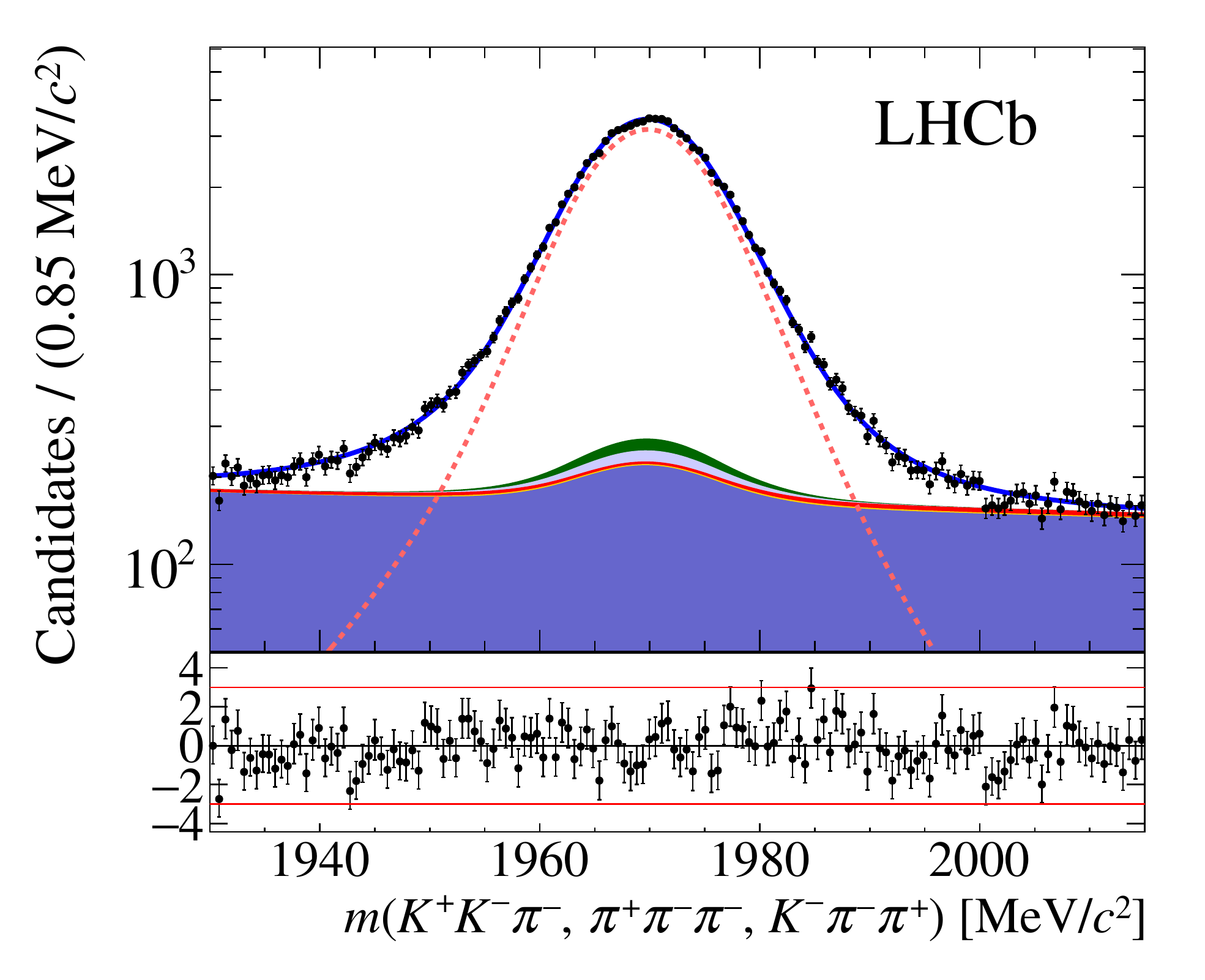}
  \includegraphics[width=.485\textwidth]{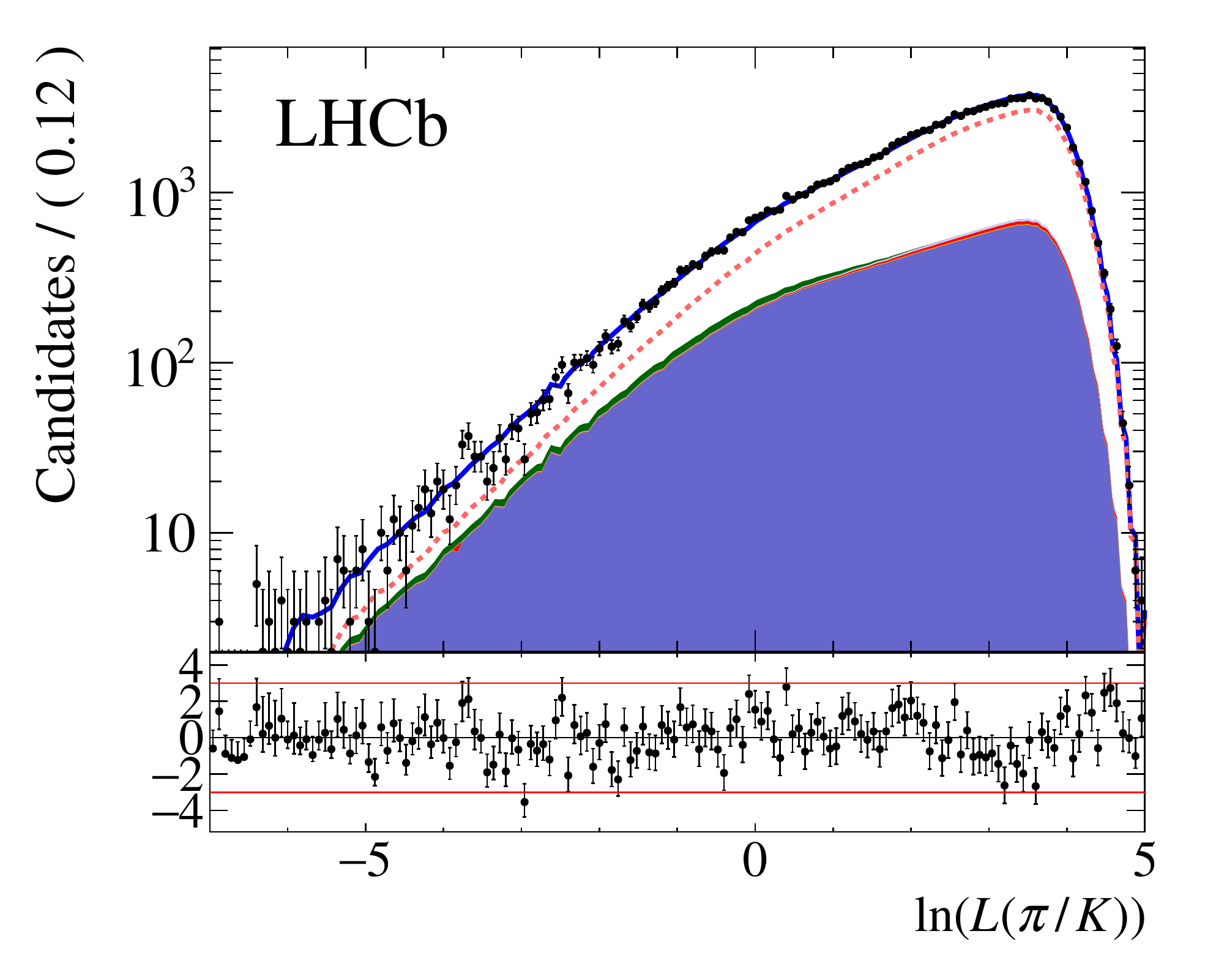}
  \caption{Distributions of the (upper left) \Bs and (upper right) \Dsm invariant masses for \BsDsPi final states, and (bottom) 
of the logarithm of the companion track PID log-likelihood, $\ln(L(\pi/K))$.
In each plot, the contributions from all \Dsm final states are combined. The solid blue curve is the total result of the simultaneous fit. The dotted red curve shows the \BsDsPi signal and the fully coloured stacked histograms show the different background contributions. Normalised residuals are shown underneath all distributions.}
  \label{fig:massfit-BsDsPi-All}
\end{figure}
The functional form of the combinatorial background is taken from the \Bs invariant mass sideband (above $5800\mevcc$),
with all parameters left free to vary in the multivariate fit.
It is parametrised separately for each \Dsm mode either by an exponential function
or by the sum of an exponential function and a constant offset.
The shapes of the fully or partially reconstructed backgrounds are fixed from
simulated events, corrected to reproduce the PID efficiency and kinematics in data,
using a nonparametric kernel estimation method (KEYS)~\cite{Cranmer:2000du}. 
An exception is background due to \Bd mesons decaying to the same final state as signal,
which is parametrised by the signal PDF shifted by the known \Bd--\Bs mass difference.

The \Dsm invariant mass is also described by a sum of two Crystal Ball functions with a common mean. 
The signal PDFs are obtained from simulation separately for each \Dsm decay mode.
As for the \Bs invariant mass signal shape, only the common mean and the width scale factor
are left free in the fits; the \Bs and \Dsm scale factors are different.
The combinatorial background consists of random combinations of tracks 
that do not originate from a \Dsm meson decay 
and backgrounds that contain a true \Dsm decay combined with a random companion track. 
Its shape  is parametrised, separately for each \Dsm decay mode, by a combination of an exponential function and the corresponding \Dsm signal PDF.
The fully and partially reconstructed backgrounds that contain a  
correctly reconstructed \Dsm candidate (\BsDsK and \BdDsPi as backgrounds in the
\BsDsPi fit; \BdDsK, \BsDsstPi, \BsDsRho and \BsDsPi as backgrounds in the \BsDsK fit)
are assumed to have the same \Dsm invariant mass distribution as the signal. 
The shapes of the other backgrounds are KEYS templates taken from simulation. 

The PDFs describing the \pidk distributions of pions, kaons and protons are obtained from
dedicated data-driven calibration samples \cite{LHCb-PROC-2011-008}.
\begin{figure}[htbp]
  \centering
  \includegraphics[width=.485\textwidth]{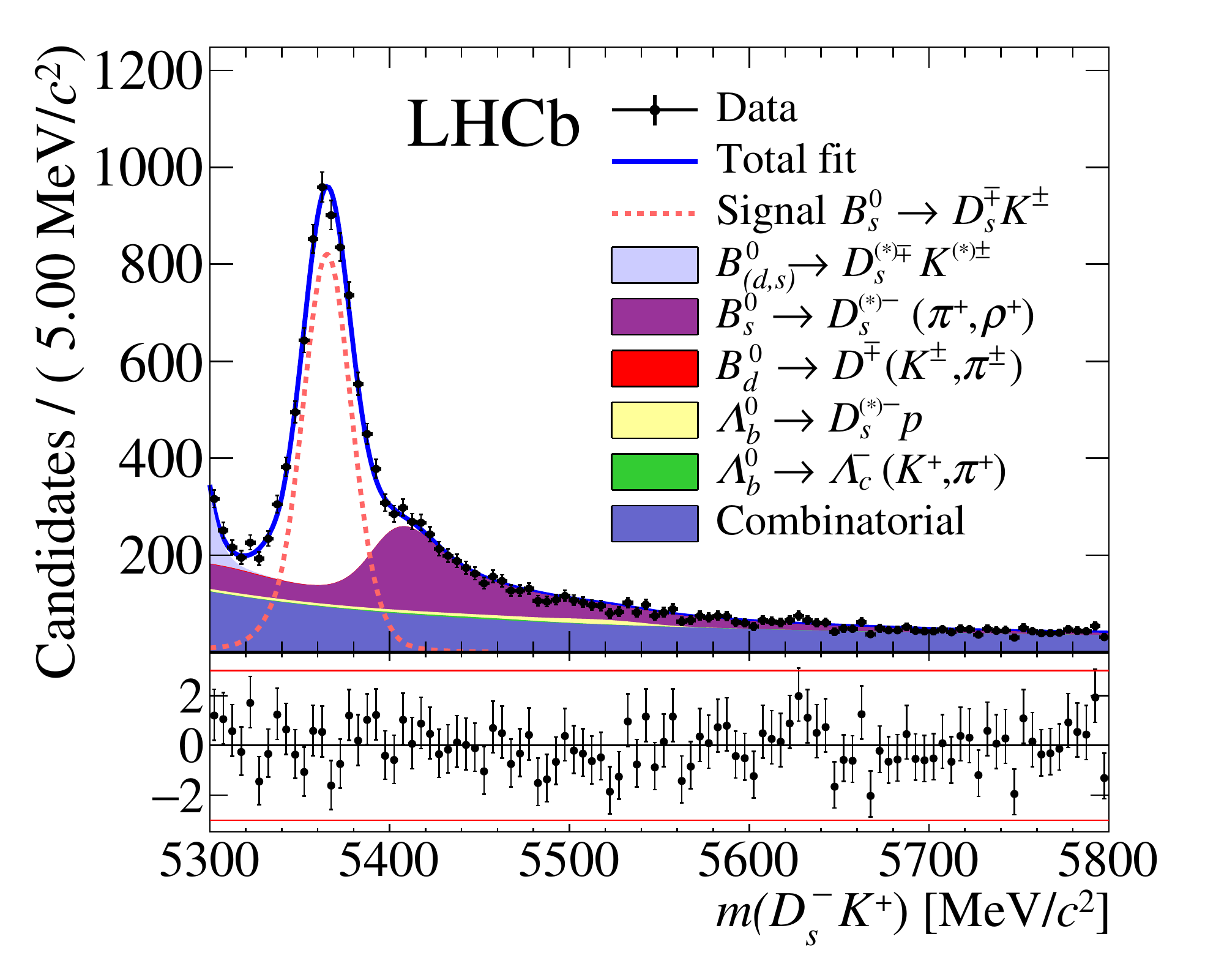}
  \includegraphics[width=.485\textwidth]{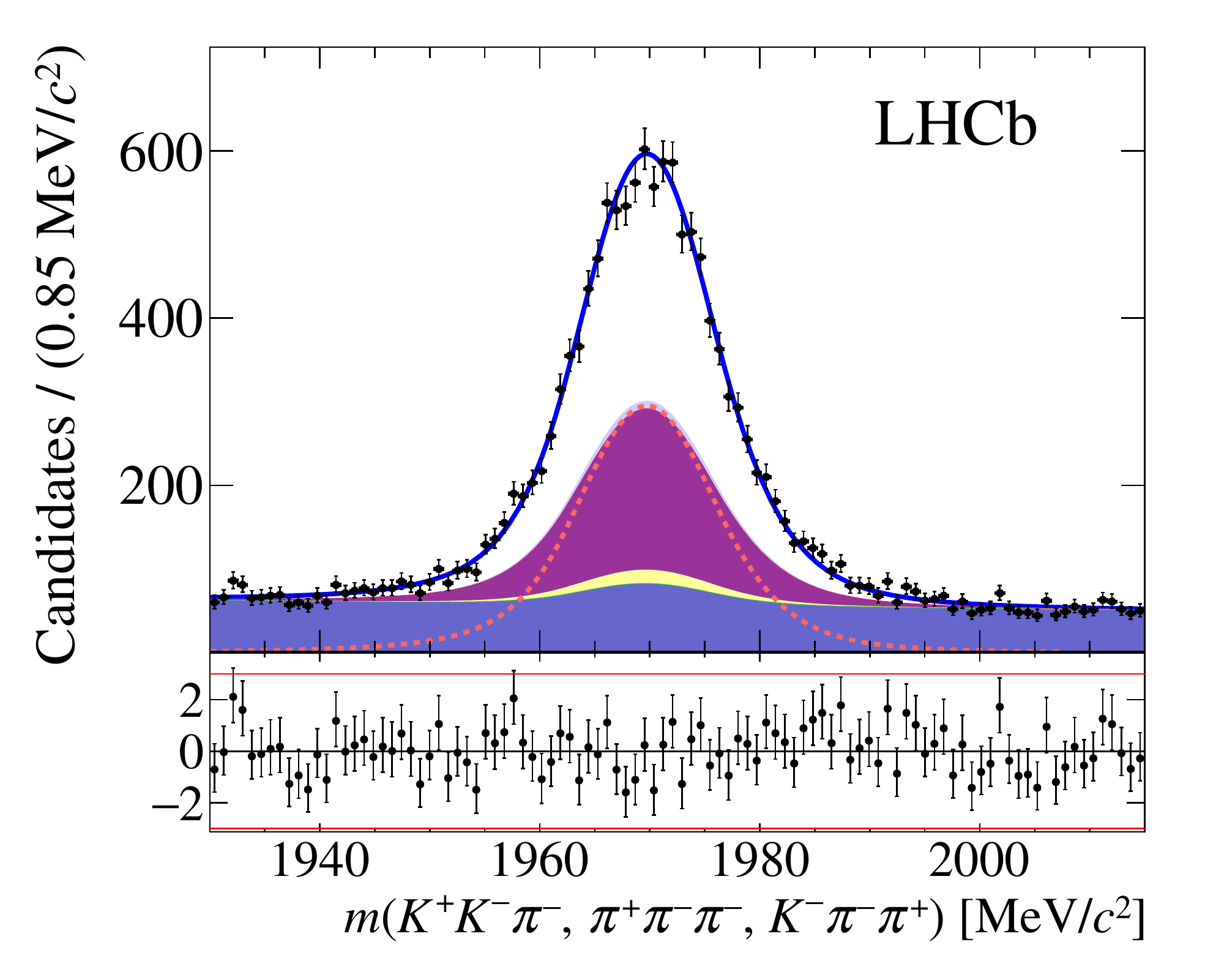}
  \includegraphics[width=.485\textwidth]{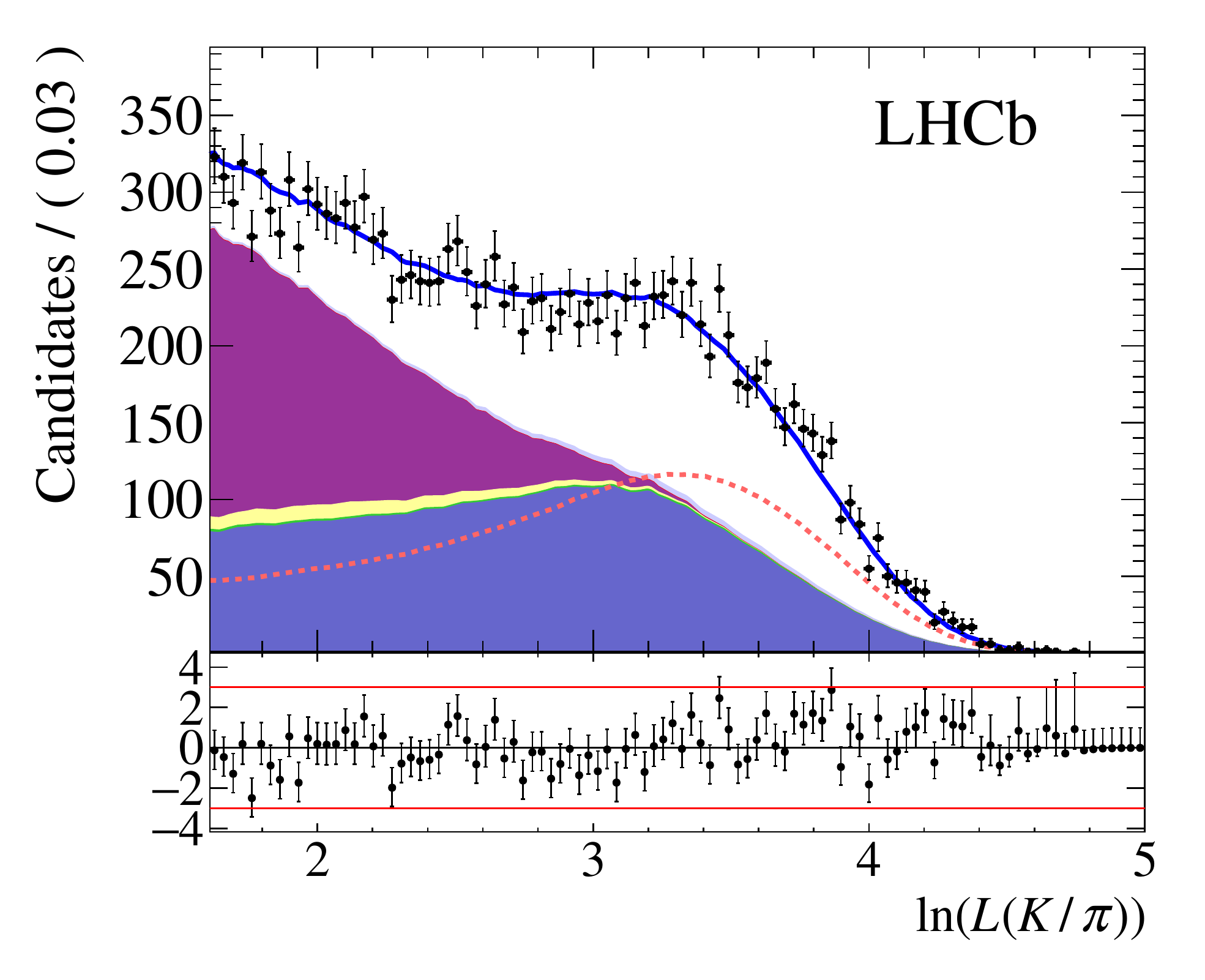}
  \caption{Distributions of the (upper left) \Bs and (upper right) \Dsm invariant masses for \BsDsK final states, and (bottom) 
of the logarithm of the companion track PID log-likelihood, $\ln(L(K/\pi))$.
In each plot, the contributions from all \Dsm final states are combined. The solid blue curve is the total result of the simultaneous fit. The dotted red curve shows the \BsDsPi signal and the fully coloured stacked histograms show the different background contributions. Normalised residuals are shown underneath all distributions.}
  \label{fig:massfit-BsDsK-All}
\end{figure}
The \pidk shape of the companion track for the signal is obtained separately for each \Dsm decay mode to account
for small kinematic differences between them. For the combinatorial background, the \pidk PDF is determined
from a mixture of pion, proton, and kaon contributions, and its normalisation
is left free in the multivariate fit. For fully or
partially reconstructed backgrounds the \pidk PDF is obtained by weighting the PID calibration samples
to match the event distributions of simulated events, separately for each background type.

The multivariate fit is performed simultaneously to the different \Dsm decay modes.
For each \Dsm decay mode the PDF is built from the sum of signal and background contributions. 
Each contribution consists of the product of three PDFs corresponding to the \Bs  and  \Ds invariant masses and \pidk, 
since their correlations are measured to be small in simulation.  
A systematic uncertainty is assigned to account for the impact of residual correlations.

Almost all background yields are left free to vary in the fit, except those that have an expected contribution below 2\% of the signal yield, namely:
 \BdDK, \BdDPi, \LbLcK, and \LbLcPi for the \BsDsK fit, and \BdDPi, \LbLcPi, and \BsDsK for the \BsDsPi fit.
Such background yields are fixed from known branching fractions and relative efficiencies measured using simulation.

The multivariate fit results in total signal yields of $96\ 942\pm 345$ and $5955 \pm 90$ \BsDsPi and \BsDsK signal candidates, respectively.
Signal yields are increased by a factor of 3.4 with respect to
the previous measurement~\cite{LHCb-PAPER-2014-038}, while the combinatorial background 
contribution is significantly reduced.
The multivariate fit is found to be unbiased using large samples of data-like pseudoexperiments.
The results of the multivariate fit are shown in Figs.~\ref{fig:massfit-BsDsPi-All} and~\ref{fig:massfit-BsDsK-All} 
for the \BsDsPi and the \BsDsK candidates, respectively, summed over all \Dsm decay modes.

\section{Flavour tagging}
\label{sec:tagging}

The identification of the \Bs initial flavour is
performed by means of different flavour-tagging algorithms. 
The same-side kaon (SS) tagger \cite{LHCb-PAPER-2015-056} searches for an additional charged 
kaon accompanying the fragmentation of the signal \Bs or \Bsb. 
The opposite-side (OS) taggers \cite{LHCb-PAPER-2011-027} exploit the pair-wise production 
of $b$ quarks that leads to a second \bquark-hadron alongside the signal $B^0_s$. 
The flavour of the nonsignal $b$ hadron is determined 
using the charge of the lepton ($\mu$, $e$) produced in semileptonic $B$ decays, or that of the kaon
from the $b\to c\to s$ decay chain, or the charge of the inclusive
secondary vertex reconstructed from $b$-decay products. 
The different OS taggers are combined and used in this analysis.

Each of these algorithms has an intrinsic mistag rate $\mistag=(\textrm{wrong
tags})/(\textrm{all tags})$, for example due to selecting tracks from the underlying event,
 particle misidentifications, or flavour oscillations of neutral
$B$ mesons on the opposite side. 
The statistical precision of the \CP-violating parameters that 
can be measured in \BsDsK decays scales as the inverse square root of the
effective tagging efficiency $\varepsilon_{\rm eff} =  \etag ( 1 - 2 \mistag)^2$, 
where $\etag$ is the fraction of signal having a tagging decision. 

The tagging algorithms are optimised to obtain the highest possible value of \effeff on data.
For each signal \Bs candidate the tagging algorithms predict
a mistag probability $\eta$ through the combination of various inputs, such
as kinematic variables of tagging particles and of the \Bs candidate, into
neural networks. The neural networks are trained on simulated samples of
\BsDsPi decays for the SS tagger and on data samples of $B^{+}\to J/\psi K^{+}$ decays for the
OS taggers.
For each tagger, the predicted mistag probability, $\eta$, is calibrated to match 
the mistag rate, \mistag, measured in data by using flavour-specific decays.
A linear model is used as a calibration function,
\begin{align}
  \label{eq:tagging-calibration}
  \mistag(\eta) = p_0 + p_1 \, (\eta - \langle\eta\rangle) ~,
\end{align}
where the values of the parameters $p_0$ and $p_1$ 
are measured using the \BsDsPi decay mode 
and $\langle\eta\rangle$ is fixed
to the mean of the estimated mistag probability $\eta$.
For a perfectly calibrated tagger one expects $p_1=1$ and $p_0 = \langle\eta\rangle$.
The tagging calibration parameters depend on the $B_{s}^{0}$ initial 
flavour, mainly due to the different interaction cross-sections
of $K^{+}$ and $K^{-}$ mesons with matter. 
Therefore, the measured \Bs--\Bsb tagging asymmetry is taken into account 
by introducing additional $\Delta p_{0}$, 
$\Delta p_{1}$ and $\Delta \etag$ parameters, 
which are defined as the difference of the corresponding \Bs and \Bsb values. 
The calibrated mistag is treated as a per-candidate variable, thus
adding an observable to the fit. 
The compatibility between the calibrations in \BsDsPi and \BsDsK
decays is verified using simulation.

\begin{table}[hbt]
  \caption{Calibration parameters and tagging asymmetries of the OS and SS taggers obtained from \BsDsPi decays. The first uncertainty is statistical and the second is systematic. 
    }
  \label{tab:tagging-calibration-DsPi}
  \centering
      {
  \begin{tabular}{ccccc}    \hline 
 &$\langle\eta\rangle$  & $p_0$  & $p_1$ &   $\etag$ [\%] \\     \hline
OS&  0.370 & $\phantom{-}0.3740\pm0.0061\pm0.0004$ & $1.094\pm0.063\pm0.012$ &  $\phantom{-}37.15\pm0.17$ \\
SS& 0.437 &  $\phantom{-}0.4414\pm0.0047\pm0.0002$ & $1.084\pm0.068\pm0.006$ & $\phantom{-}63.90\pm0.17$ \\  
 \hline
&--          & $\Delta p_0$             & $\Delta p_1$             &   $\Delta \etag$ [\%]  \\ \hline
OS& --         &  $\phantom{-}0.0138\pm0.0060\pm0.0001$ & $0.126\pm0.062\pm0.002$ & $-1.14\pm0.72$ \\
SS &  -- & $-0.0180\pm0.0047\pm0.0002$ & $0.134\pm0.067\pm0.002$ & $\phantom{-}0.82\pm0.72$ \\  
    \hline 
  \end{tabular}
}
\end{table}

\begin{table}[hbt!]
  \caption{Performances of the flavour tagging for \BsDsPi candidates tagged by OS only, 
 SS only and both OS and SS algorithms.}
  \label{tab:tagging-performances}
  \centering
  \begin{tabular}{ccc}
    \hline
    \BsDsPi & \etag  [\%]                       & \effeff [\%]          \\
    \hline
    OS only & $12.94 \pm 0.11$ & $1.41 \pm 0.11$ \\
    SS only & $39.70 \pm 0.16$ & $1.29 \pm 0.13$ \\
    Both OS and SS & $24.21 \pm 0.14$ & $3.10 \pm 0.18$ \\
    \hline
    Total        &  $76.85 \pm 0.24$    &   $5.80 \pm 0.25$ \\
    \hline
 \end{tabular}
\end{table}

The flavour-specific \BsDsPi decay mode is used for tagging calibration in order to  minimize the systematic uncertainties 
due to the portability of the calibration from a different control
channel to the signal one. 
The measured values of the OS and SS tagging calibration parameters and tagging asymmetries in 
the \BsDsPi sample are summarised in Table~\ref{tab:tagging-calibration-DsPi}.
They are obtained from a fit to the decay-time distribution of the \BsDsPi sample in which the background is  
statistically subtracted by weighting the candidates according to the weights computed with the multivariate fit. 
The measured effective tagging efficiency for the inclusive OS and SS taggers is approximately 3.9\% and 2.1\%, respectively.
The results of the 2011 and 2012 samples are consistent.

Systematic uncertainties on the calibration parameters have an impact on the \CP parameters and 
they are added in quadrature with the statistical uncertainties and used to define the Gaussian constraints on the calibration parameters in the \BsDsK fit.
The largest systematic effect on the tagging calibration parameters is due to the decay-time resolution model, which also affects the \BsDsK fit for \CP observables. 
In order to avoid double counting, this source of systematic uncertainty is treated separately from the other systematic sources (see Sec.~\ref{sec:systematics}).
Other relevant sources of systematic uncertainties are related to the calibration method and 
to the background description in the multivariate fit used to compute the weights for the \sfit procedure.
Uncertainties related to the decay-time acceptance and to the fixed values of $\Delta m_{s}$ and $\Delta \Gamma_{s}$ in the \sfit
procedure are found to be negligible.
The total systematic uncertainties, reported in Table~\ref{tab:tagging-calibration-DsPi}, are significantly smaller than the statistical.

The OS and SS tagging decisions and the mistag
predictions are combined  in the fit to the \BsDsK decay-time distribution
by using the same approach as described in Ref.~\cite{LHCb-PAPER-2015-004}. 
The tagging performances for the OS and SS combination measured in the \BsDsPi channel are reported in Table~\ref{tab:tagging-performances}. 
Three categories of tagged events are considered: OS only, SS only and both OS and SS.
The estimated value of the effective tagging efficiency \effeff for the \BsDsK decay mode is ($5.7 \pm 0.3$)\%, consistent with the value obtained for \BsDsPi decays, as expected.

\section{Decay-time resolution}
\label{sec:timeres}
Due to the fast \Bs--\Bsb oscillations, the \CP-violation parameters related to the amplitudes of the sine and cosine terms are highly correlated to the decay-time resolution model.
The signal decay-time PDF is convolved with a Gaussian resolution function that has a different width
for each candidate, making use of the per-candidate decay-time uncertainty
estimated from the kinematic fit of the \Bs vertex.

\begin{figure}[!tb]
  \centering
  \includegraphics[width=.54\textwidth]{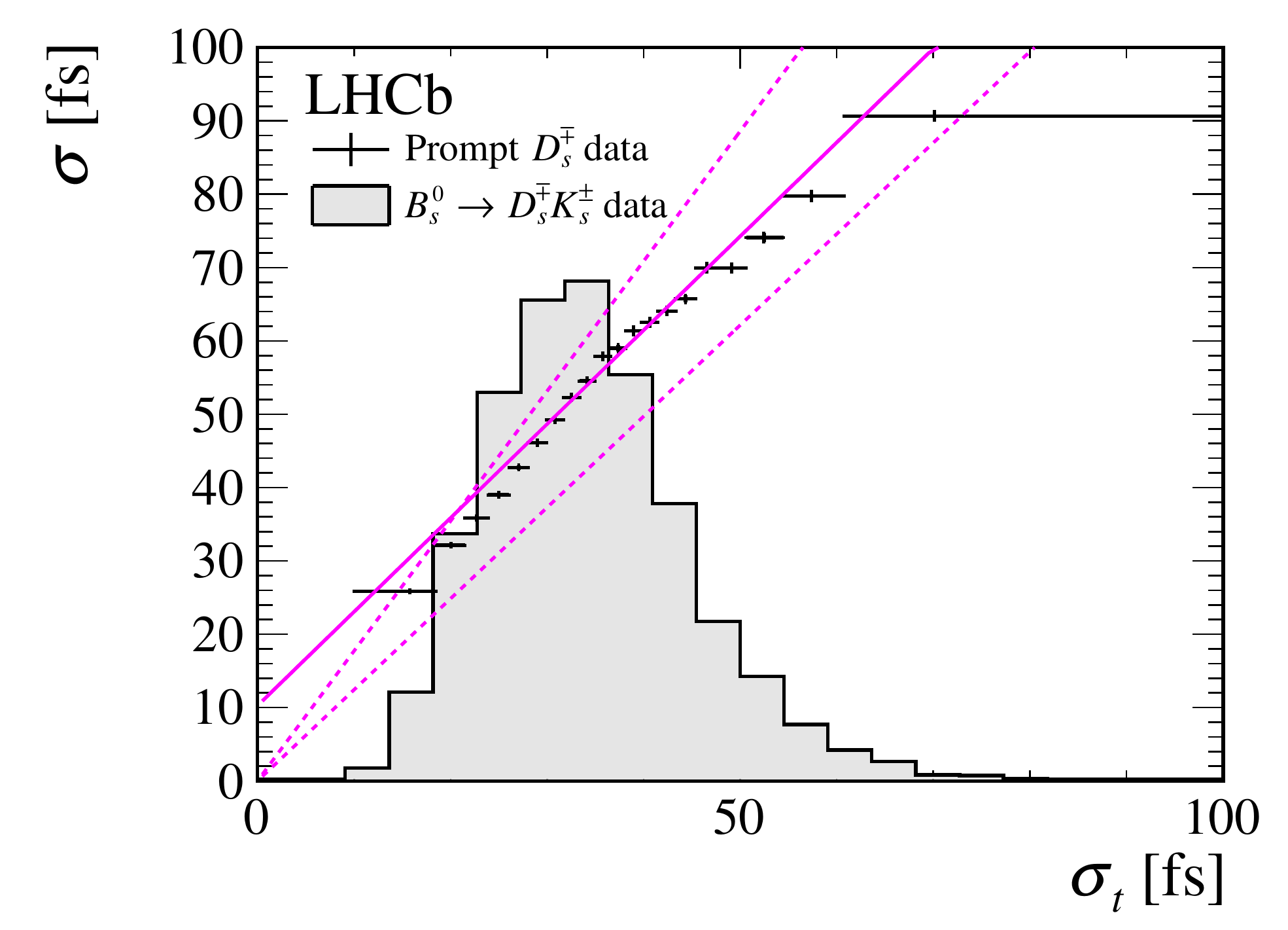} 
  \caption{Data points show the measured resolution $\sigma$ as a function of the
    per-candidate uncertainty $\sigma_t$ for prompt $D_s^{\mp}$ candidates combined with a random track.
    The dashed lines indicate the values used to determine the
    systematic uncertainties on this method. The solid line shows the linear fit to the
    data as discussed in the text. The histogram overlaid
    is the distribution of the per-candidate decay-time uncertainty for 
    \BsDsK candidates.}
  \label{fig:promptDs-resvserror}
\end{figure}

From the comparison to the measured decay-time resolution, a correction to the per-candidate decay-time uncertainty $\sigma_{t}$ is determined.
This
calibration is performed from a sample of ``fake \Bs'' candidates with a known lifetime of zero obtained from the combination of prompt \Dsm mesons with a random track that originated from the PV. The
spread of the observed decay times follows the shape of a double Gaussian
distribution, where only the negative decay times are used to determine the resolution, to avoid biases in the determination of the decay-time resolution due to long-lived backgrounds.
The resulting two widths are combined to calculate the corresponding dilution: 
\begin{align*}
D = f_1 e^{-\sigma_1^2 \dms^2/2} + (1-f_1) e^{-\sigma_2^2 \dms^2/2}, 
\end{align*}
where
$\sigma_{1,2}$ are the widths, and $f_1$ and $(1-f_1)$ are the fractions of the two
Gaussian components. The dilution, which represents the amplitude damping  of the decay-time distribution, is used to obtain the effective decay-time resolution $\sigma = \sqrt{(-2 \pi/ \Delta m_s^2) \ln(D)}$.
The effective decay-time resolution depends on the
per-candidate decay-time uncertainty as 
$\sigma(\sigma_{t}) = 1.28 \,\sigma_{t} + 10.3~\mathrm{fs}$, and is shown in Fig.~\ref{fig:promptDs-resvserror}.  
The uncertainty on the decay-time resolution is dominated by
the uncertainty on the modelling of the observed decay times of the ``fake \Bs''
candidates. Modelling the spread by a single Gaussian distribution or by taking only the 
central Gaussian from the double Gaussian fit, results in the correction factors 
$\sigma(\sigma_{t}) = 1.77 \,\sigma_{t}$ and 
$\sigma(\sigma_{t}) = 1.24 \,\sigma_{t}$, respectively, 
which are used to estimate the systematic uncertainty on the measured \CP parameters.

The assumption that the measured decay-time resolution on ``fake \Bs'' candidates can be used for
true \Bs candidates is justified, as the measured decay-time resolution does not
significantly depend on the transverse momentum of the companion particle, which
is the main kinematic difference between the samples.  In addition, simulation
shows that the ``fake \Bs'' and signal \Bs samples require compatible correction factors, 
varying in the range $[1.19,1.27]$.

\section{Decay-time acceptance}
\label{sec:timeacc}
\begin{figure}[hbt!]
 \centering
 \includegraphics[width=.60\textwidth]{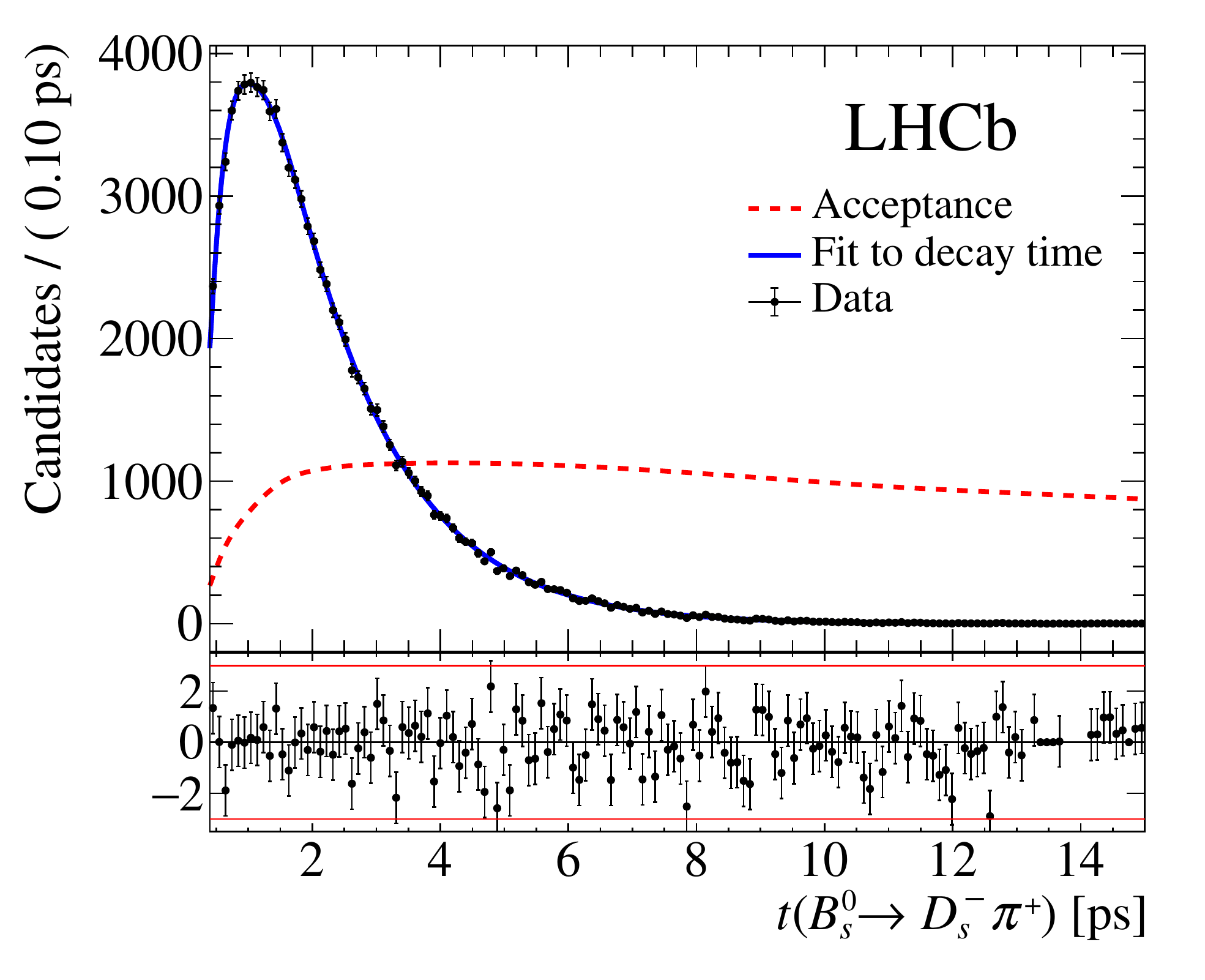} 
 \caption{Decay time distribution of \BsDsPi candidates obtained by the \emph{sPlot} technique. The solid blue curve is the result of the \sfit procedure and the dashed 
red curve shows the measured decay-time acceptance in arbitrary units. Normalised residuals are shown underneath.}
\label{fig:timesfit_bsdspi}
\end{figure}

The decay-time acceptance of \BsDsK candidates 
is strongly correlated with the \CP parameters, in particular with \Dpar and \Dbpar. 
However, in the case of the flavour-specific \BsDsPi decays, the acceptance can be measured 
by fixing \gs and floating the acceptance parameters. 
The decay-time acceptance in the \BsDsK fit is fixed
to that found in the fit to \BsDsPi data, corrected by the acceptance ratio in the two channels obtained from simulation,
which is weighted as described in Sec.~\ref{sec:mdfit}.
In all cases, the acceptance is described using 
segments of cubic b-splines, which are implemented in an analytic way in the decay-time fit \cite{Karbach:ComplxErrFunc}. 
The spline boundaries, knots, are chosen in order to model reliably
the features of the acceptance shape, and are placed at $0.5$, $1.0$, $1.5$, $2.0$, $3.0$ and $12.0\ps$. 
In the \sfit procedure applied to the sample of \BsDsPi candidates, the \CP-violation parameter \Cpar is fixed to unity with $\Cpar = -\Cbpar$, while \Spar, \Sbpar, \Dpar, and \Dbpar are all fixed
to zero. The spline parameters and \dms are free to vary. 
The result of the \sfit procedure applied to the \BsDsPi candidates is shown in Fig.~\ref{fig:timesfit_bsdspi}.

Extensive studies with simulation have been performed and confirm the validity of the method.
An alternative analytical decay-time acceptance parametrisation has been considered,
and is in good agreement with the nominal spline description. 
Finally, doubling the number of knots results in negligible changes in the fit result. 

\boldmath
\section{Decay-time fit to \BsDsK}
\label{sec:timefit}
\unboldmath

In the \sfit procedure applied to the \BsDsK candidates, the following parameters 
\begin{equation}
\begin{aligned}
\Delta m_s &= (17.757 \pm 0.021)\invps\,, \\ 
\Gamma_s &= (0.6643 \pm 0.0020)\invps\,,  \\
\Delta\Gamma_s &= (0.083 \pm 0.006)\invps\,,  \\
\rho(\Gs,\DGs)&=-0.239\,, \\
A_{\textrm{prod}} &= (1.1 \pm 2.7)\%,   \\
A_{\textrm{det}} &= (1 \pm 1)\%  
\end{aligned}
\end{equation}
are fixed to their central values. 
The values of \Bs oscillation frequency and production asymmetry, $A_{\rm prod}$, are based on LHCb measurements ~\cite{LHCb-PAPER-2013-006,LHCb-PAPER-2014-042}.
The \Bs decay width, $\Gamma_s$, the decay-width difference, $\Delta \Gamma_s$, and their correlation, $\rho(\Gs,\DGs)$, correspond to the HFLAV~\cite{HFAGSpring16} world average.
An estimate of the detection asymmetry $A_{\textrm{det}}$ based on Ref.~\cite{LHCb-PAPER-2014-013} is considered. 
The production asymmetry is defined as $A_{\textrm{prod}} \equiv [\sigma(\Bsb) - \sigma(\Bs)]/[\sigma(\Bsb) + \sigma(\Bs)]$, 
where $\sigma$ denotes the production cross-section inside the LHCb acceptance. 
The detection asymmetry is defined as the difference in reconstruction efficiency between the 
$\Dsm K^+$ and the $\Dsp K^-$ final states. 
The detection and the production asymmetries contribute to the PDF with factors of $(1 \pm A_{\textrm{prod}})$ and 
$(1 \pm A_{\textrm{det}})$, depending on the tagged initial state and the reconstructed final state, respectively. 
The tagging calibration parameters and asymmetries are allowed to float within Gaussian constraints based on their statistical and systematic uncertainties given in Sec.~\ref{sec:tagging}. 
The decay-time PDF is convolved with a single Gaussian
representing the per-candidate decay-time resolution, and multiplied by the decay-time acceptance
described in Sec.~\ref{sec:timeres} and Sec.~\ref{sec:timeacc}, respectively.

\begin{figure}[h]
 \centering
 \includegraphics[width=.60\textwidth]{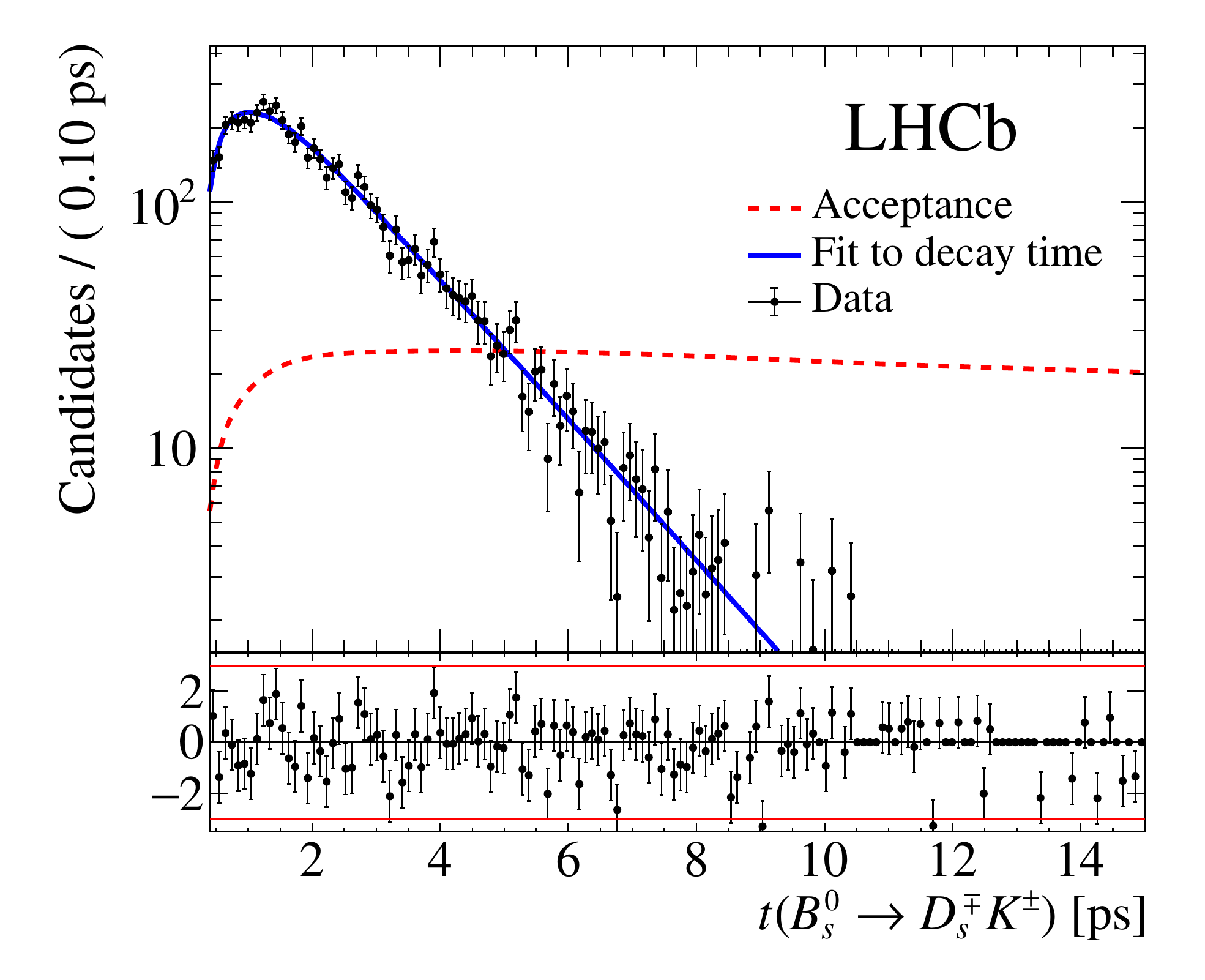}
 \includegraphics[width=.98\textwidth]{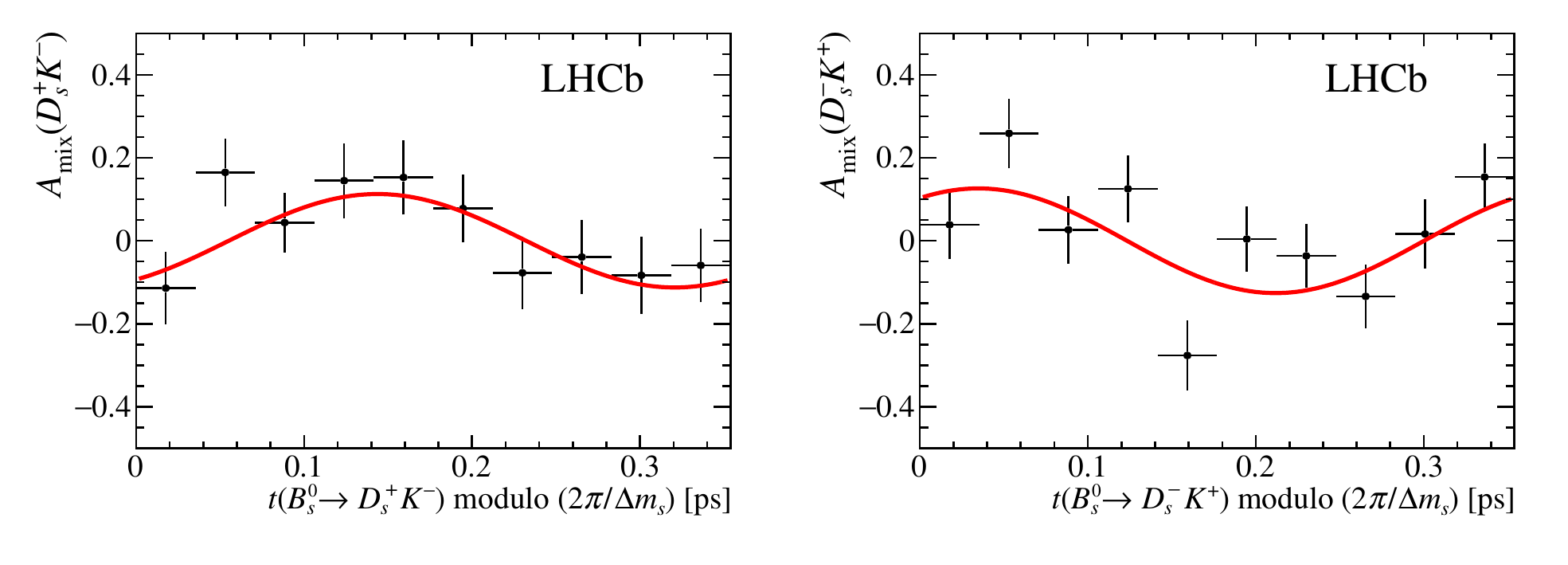}
	\caption{The (top) decay-time distribution of \BsDsK candidates obtained by the \emph{sPlot} technique.
The solid blue curve is the result of the \sfit procedure and the dashed red curve shows the decay-time acceptance in arbitrary units, obtained from the \sfit procedure applied to the \BsDsPi candidates and corrected for the ratio of decay-time acceptances of \BsDsK and \BsDsPi from simulation. Normalised residuals are shown underneath.
     The \CP-asymmetry plots for (bottom left) the $\Dsp K^-$ final state and (bottom right) the $\Dsm K^+$ final state, folded into one mixing period $2\pi/\dms$, are also shown.}
 \label{fig:timesfit_bsdsk}
\end{figure}

The measured \CP-violating parameters are given in Table~\ref{tab:timefit_bsdsk}, and
the correlations of their statistical uncertainties
are given in Table~\ref{tab:timefit_bsdskcorr}. The fit to the decay-time distribution is shown in Fig.~\ref{fig:timesfit_bsdsk}. 
together with the two decay-time-dependent asymmetries, $A_{\rm mix}(D^{+}_{s}K^{-})$ and $A_{\rm mix}(D^{-}_{s}K^{+})$, 
that are defined as the difference of the decay rates (see Eqs.~\ref{eq:decay_rates_1} and~\ref{eq:decay_rates_2})
of the tagged candidates. The asymmetries are obtained by folding the decay time in one mixing period $2\pi/\dms$.
The central values of the \CP parameters measured by the fit 
are used to determine the plotted asymmetries.

\begin{table}[t]
\centering
\caption{Values of the \CP-violation parameters obtained from the fit  to the decay-time distribution of \BsDsK decays.
The first uncertainty is statistical and the second is systematic. }
\label{tab:timefit_bsdsk}
\begin{tabular}{lc}
  \hline
  Parameter     & Value \\
  \hline
  $\quad$ \Cpar         & $\phantom{+}0.730   \pm  0.142 \pm 0.045$  \\ 
  $\quad$ \Dpar         & $\phantom{+}0.387  \pm  0.277 \pm 0.153$  \\
  $\quad$ \Dbpar        & $\phantom{+}0.308  \pm  0.275 \pm 0.152$  \\
  $\quad$ \Spar         & $-0.519   \pm  0.202 \pm 0.070$  \\
  $\quad$ \Sbpar        & $-0.489  \pm  0.196 \pm 0.068$  \\
  \hline
\end{tabular}
\vspace{-3mm}
\end{table}
\begin{table}[t]
\centering
\caption{Statistical correlation matrix of the \CP parameters.
Other fit parameters have negligible correlations with the \CP parameters.
}
\label{tab:timefit_bsdskcorr}
\begin{tabular}{lrrrrr}
  \hline
\rule{0pt}{2.5ex}    Parameter        & \Cpar   & \Dpar   & \Dbpar  & \Spar   & \Sbpar\\ 
  \hline
 $\quad$ \Cpar     & $\phantom{+}1 $  & $\phantom{+}0.092$              &$\phantom{+}0.078 $             & $\phantom{+}0.008$             & $-0.057 $  \\
 $\quad$ \Dpar     &  & $\phantom{+}1 $  &$\phantom{+} 0.513 $  & $-0.083$             & $-0.004$  \\
 $\quad$ \Dbpar    &  &  &$\phantom{+} 1 $  & $-0.042$             & $-0.003$  \\
 $\quad$ \Spar     &  &  &   & $\phantom{+}1 $  & $\phantom{+}0.001$  \\
 $\quad$ \Sbpar    &  &  &   &  & $\phantom{+}1 $  \\
 \hline
\end{tabular}
\end{table}
\section{Systematic uncertainties}
\label{sec:systematics}
Systematic uncertainties arise from
the fixed parameters \dms,\ \gs,\ \dgs,\  the detection $A_{\rm det}$ and tagging efficiency $\Delta \etag$ asymmetries,
and from the limited knowledge of the decay-time resolution and acceptance.
In addition, the impact of neglecting correlations among the observables for background candidates is estimated. 
Table~\ref{tab:TotalSystErrSfit} summarises the different contributions to the systematic uncertainties, which are detailed below.

The systematic uncertainties are estimated
using large sets of pseudoexperiments, in which the relevant parameters are
varied. The pseudoexperiments are generated 
with central values of the \CP parameters reported in Sec.~\ref{sec:timefit}.
They are subsequently processed by the same fit procedure applied to data. 
The fitted values are compared between
the nominal fit, where all fixed parameters are kept at their nominal values,
and the systematic fit, where each parameter is varied according to its
uncertainty. A distribution is formed by normalising the resulting
differences to the uncertainties measured in the nominal fit, and the 
mean and width of this distribution are added in quadrature and assigned as the
systematic uncertainty.

The systematic uncertainty related to the decay-time resolution model, together with its impact on the flavour tagging, is evaluated
by fitting the \BsDsK pseudoexperiments using the two alternative decay-time resolution models and their corresponding tagging calibration parameters.
The latter are obtained with \BsDsPi pseudoexperiments that were generated with the nominal decay-time resolution, 
but fitted with the two alternative decay-time resolution models. 
\begin{table}[htb!]
\centering
\caption{Systematic uncertainties on the \CP parameters, relative to the statistical uncertainties.}
\label{tab:TotalSystErrSfit}
\begin{tabular}{lccccc}
\hline
\rule{0pt}{2.5ex}  Source                                   & \Cpar  & \Dpar   & \Dbpar & \Spar  & \Sbpar \\
\hline
Detection asymmetry                 & 0.02  & 0.28  & 0.29  & 0.02  & 0.02        \\
$\Delta m_s$                        & 0.11  & 0.02  & 0.02  & 0.20  & 0.20        \\
Tagging and scale factor            & 0.18  & 0.02  & 0.02  & 0.16  & 0.18        \\ %
Tagging asymmetry                   & 0.02  & 0.00  & 0.00  & 0.02  & 0.02        \\ %
Correlation among observables       & 0.20  & 0.38  & 0.38  & 0.20  & 0.18        \\
Closure test                        & 0.13  & 0.19  & 0.19  & 0.12  & 0.12        \\
Acceptance, simulation ratio        & 0.01  & 0.10  & 0.10  & 0.01  & 0.01        \\
Acceptance data fit, \Gs, \DGs      & 0.01  & 0.18  & 0.17  & 0.00  & 0.00        \\
\hline
Total                               & 0.32  & 0.55  & 0.55  & 0.35  & 0.35        \\
\hline
\end{tabular}
\end{table}
The impact of neglecting the correlations among the observables in the background is accounted for by means of a dedicated set of pseudoexperiments
in which the correlations are included at generation and neglected in the fit. 
The correlations between \Gs, \DGs, 
and the decay-time acceptance parameters from the fit to \BsDsPi data
are accounted for by fitting pseudoexperiments, where the values of the spline coefficients, \Gs and \DGs are randomly generated
according to multidimensional correlated Gaussian distributions centred at the nominal values.
The combined correlated systematic uncertainty is listed as ``acceptance data fit, \Gs, \DGs''.
The correlations between the spline coefficients among \BsDsPi and \BsDsK simulation samples 
are accounted for
by fitting pseudoexperiments with the parameters randomly generated as in the previous case,
and the corresponding systematic uncertainty is listed as ``acceptance, simulation ratio''.

\begin{table}[htb!]
\centering
\caption{Correlation matrix of the total systematic uncertainties of the \CP parameters.}
\label{tab:TotalSystCorSfit}
\begin{tabular}{crrrrr}
\hline
\rule{0pt}{2.5ex}    Parameter & \Cpar  & \Dpar   & \Dbpar & \Spar  & \Sbpar \\
\hline
 $\quad$ \Cpar     & $\phantom{+}1 $  & $\phantom{+}0.05$              &$\phantom{+}0.03 $             & $\phantom{+}0.03$             & $-0.01$  \\
 $\quad$ \Dpar     &  & $\phantom{+}1 $  &$\phantom{+} 0.42 $  & $0.02$             & $0.02$  \\
 $\quad$ \Dbpar    &  &  &$\phantom{+} 1 $  & $0.03$             & $0.03$  \\
 $\quad$ \Spar     &  &  &   & $\phantom{+}1 $  & $\phantom{+}0.01$  \\
 $\quad$ \Sbpar    &  &  &   &  & $\phantom{+}1 $  \\
\hline
\end{tabular}
\end{table}

The nominal result is cross-checked by splitting the sample into subsets according
to the two magnet polarities, the year of data taking, the \Bs momentum, and the BDTG response. 
No dependencies are observed. 
In particular, the compatibility of the 1$\invfb$ and the 2$\invfb$ subsamples is at the level of 1 $\sigma$, where $\sigma$ is the standard deviation.
A closure test using the high-statistics fully simulated signal candidates provides an estimate of the intrinsic uncertainty related to the fit procedure. 
No bias is found and only the fit uncertainty is considered as a systematic uncertainty. 
The systematic effects due to the background subtraction in the \sfit procedure are checked. 
Therefore, the nominal fitting procedure is applied to a mixture of the signal and the \BsDsPi simulation samples as well as combinatorial background data. 
The result is consistent with the values found by the fit to the signal only, as a consequence, no additional uncertainties are considered. 

The resulting systematic uncertainties are shown in Table~\ref{tab:TotalSystErrSfit} relative to the
corresponding statistical uncertainties. 
The total systematic correlation matrix, reported in Table~\ref{tab:TotalSystCorSfit}, 
is obtained by adding the covariance matrices corresponding to each source.

A number of other possible systematic effects  are studied, but found to be negligible. These include
production asymmetries, missing or imperfectly modelled backgrounds, 
and fixed signal-shape parameters in the multivariate fit.
Potential systematic effects due to fixed background yields are evaluated by generating
pseudoexperiments with the nominal value for these yields, and fitting back with the yields
fixed to twice or half their nominal value. No significant bias is observed and no systematic uncertainty assigned.
The decay-time fit is repeated adding one or two additional spline functions
to the decay-time acceptance description and no significant change in the fit result is observed.
The multivariate and decay-time fits are repeated randomly removing multiple candidates, 
with no significant change observed in the fit result.
No systematic uncertainty is assigned to the imperfect knowledge of the
momentum and the longitudinal dimension of the detector since both effects are
taken into account by the systematic uncertainty on \dms, 
as the world average is dominated by the LHCb measurement~\cite{LHCb-PAPER-2013-006}.

\section{Interpretation}
\label{sec:interpretation}

The measurement of the \CP parameters is used to determine
the values of \weak and, subsequently, of the angle \g.
The following likelihood is maximised, replicating the procedure described in Ref.~\cite{LHCb-PAPER-2016-032},
\begin{equation}
  \label{eq:gLikelihood}
  \mathcal L(\vec \alpha) = \exp\left( -\frac 1 2 \left( \vec A(\vec \alpha) - \vec A_{\text{obs}} \right)^T V^{-1}
  \left( \vec A(\vec \alpha) - \vec A_{\text{obs}} \right) \right)\,,
\end{equation}
where $\vec \alpha = (\g,\beta_s,\rdsk,\strong)$ is the vector of the physics
parameters, $\vec{A}(\vec{\alpha})$ is the vector of parameters expressed through
Eq.~\ref{eq:truth}, $\vec A_{\text{obs}}$ is the vector of the measured
\CP-violating parameters and $V$ is the experimental (statistical and systematic) uncertainty
covariance matrix. Confidence intervals are computed by evaluating the test
statistic
\mbox{$\Delta\chi^2 \equiv \chi^2(\vec{\alpha}'_{\min}) - \chi^2(\vec{\alpha}_{\min})$},
where $\chi^2(\vec{\alpha}) = -2 \ln \mathcal{L}(\vec{\alpha})$,
following Ref.~\cite{LHCb-PAPER-2013-020}. Here, $\vec{\alpha}_{\min}$
denotes the global maximum of Eq.~\ref{eq:gLikelihood}, and $\vec{\alpha}'_{\min}$
is the conditional maximum when the parameter of interest is fixed to the tested value.

\begin{figure}
  \centering
  \includegraphics[width=.48\textwidth]{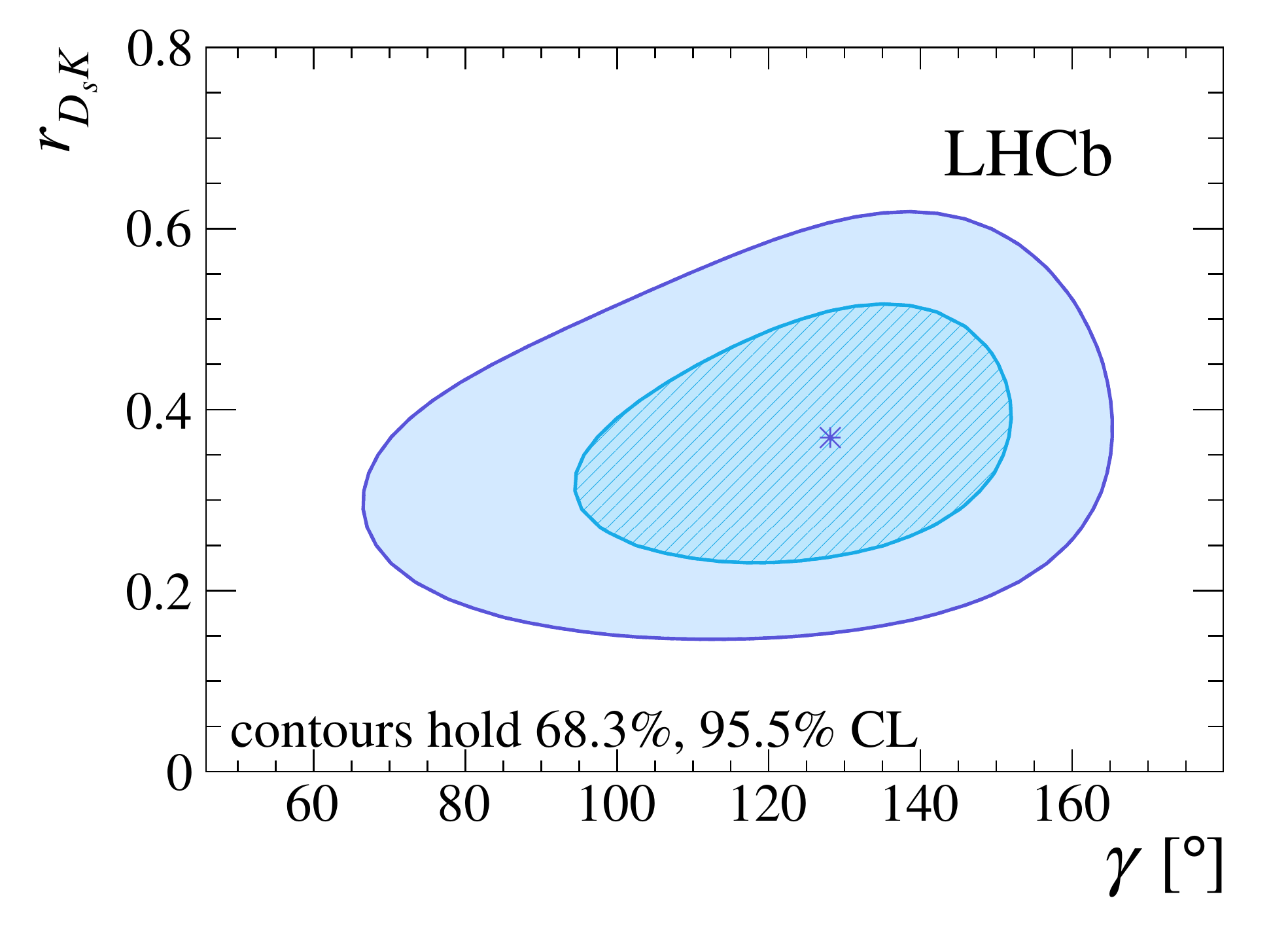}
  \includegraphics[width=.48\textwidth]{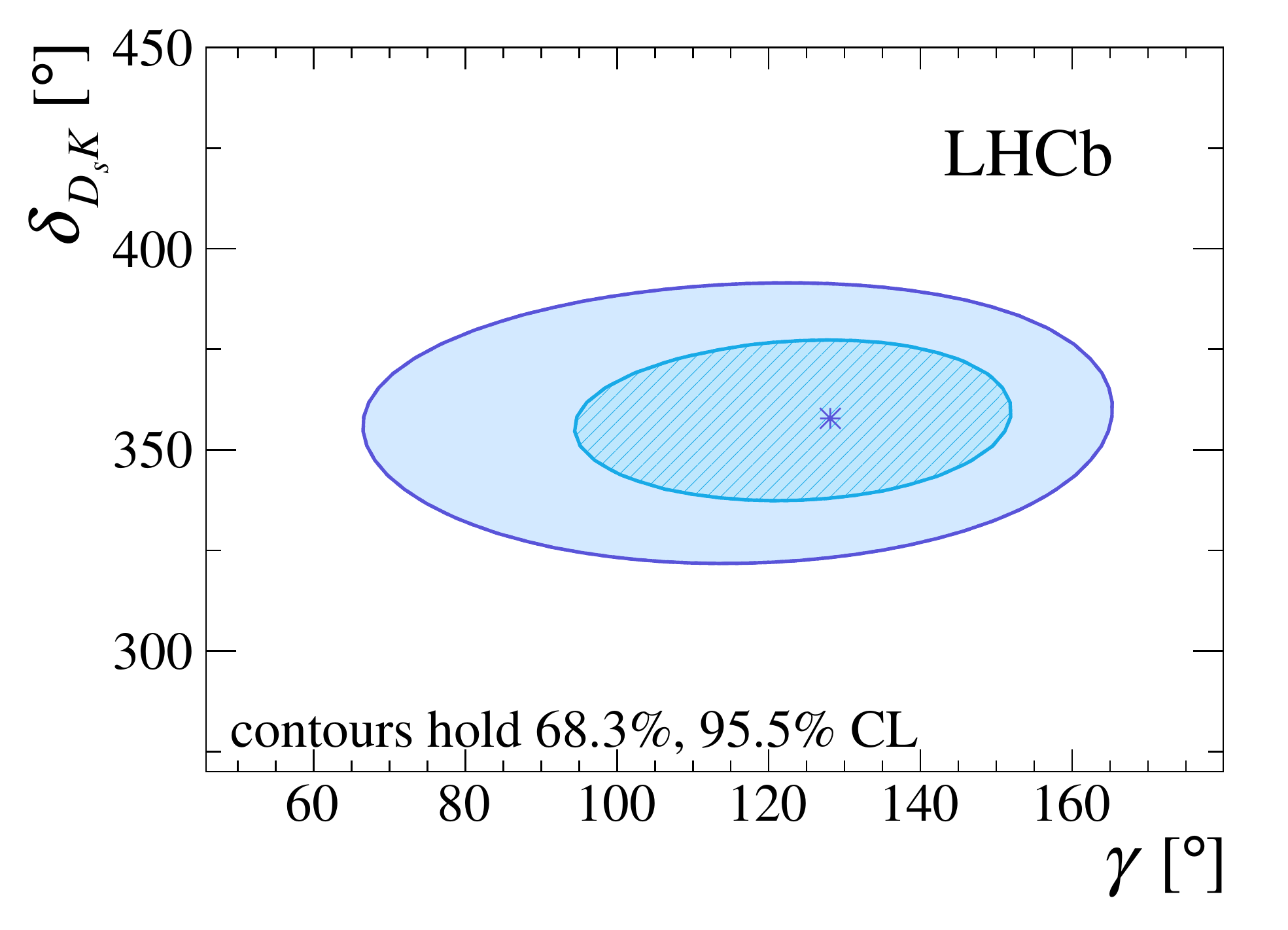} \\
  \includegraphics[width=.48\textwidth]{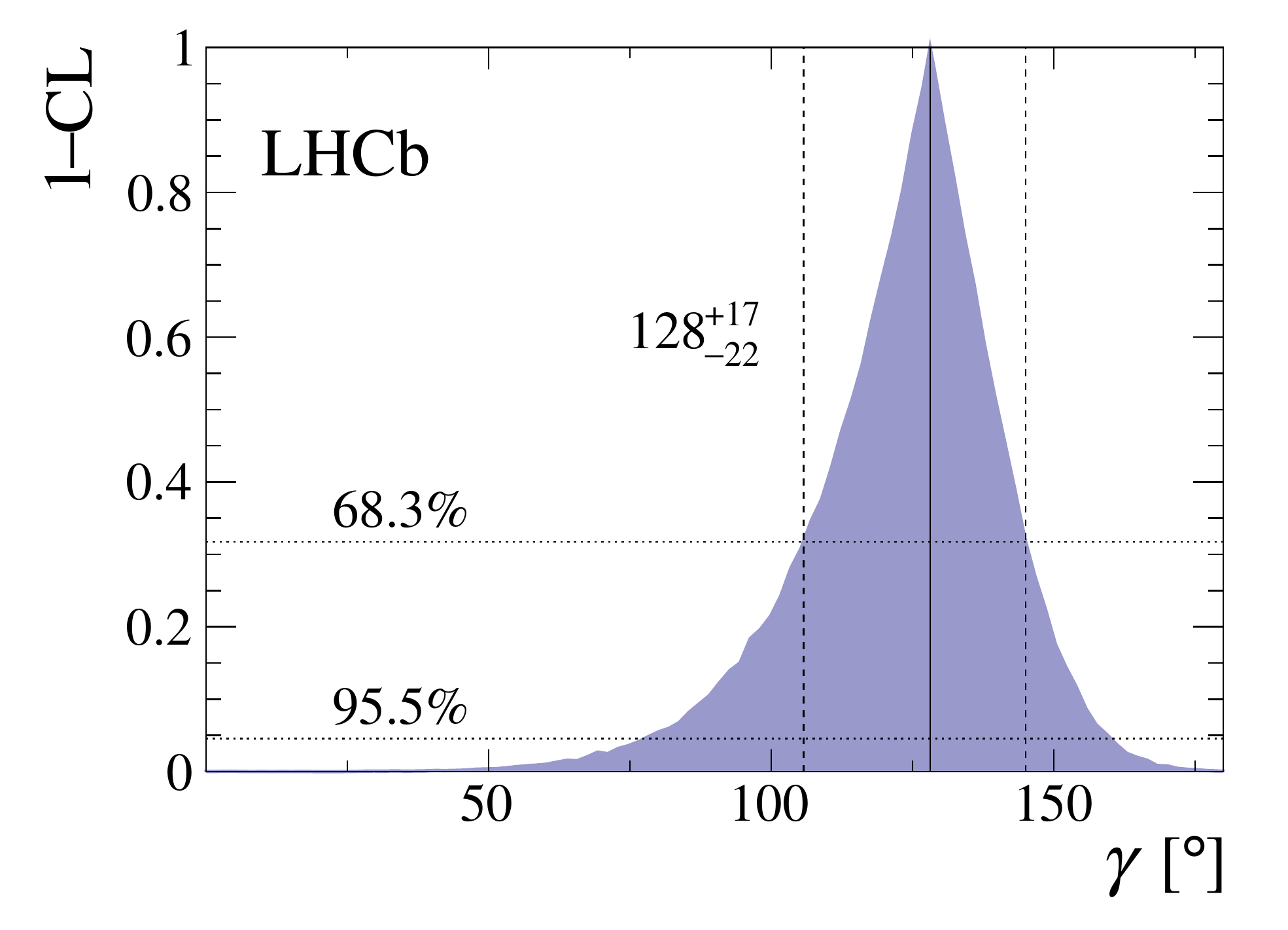} 
  \includegraphics[width=.48\textwidth]{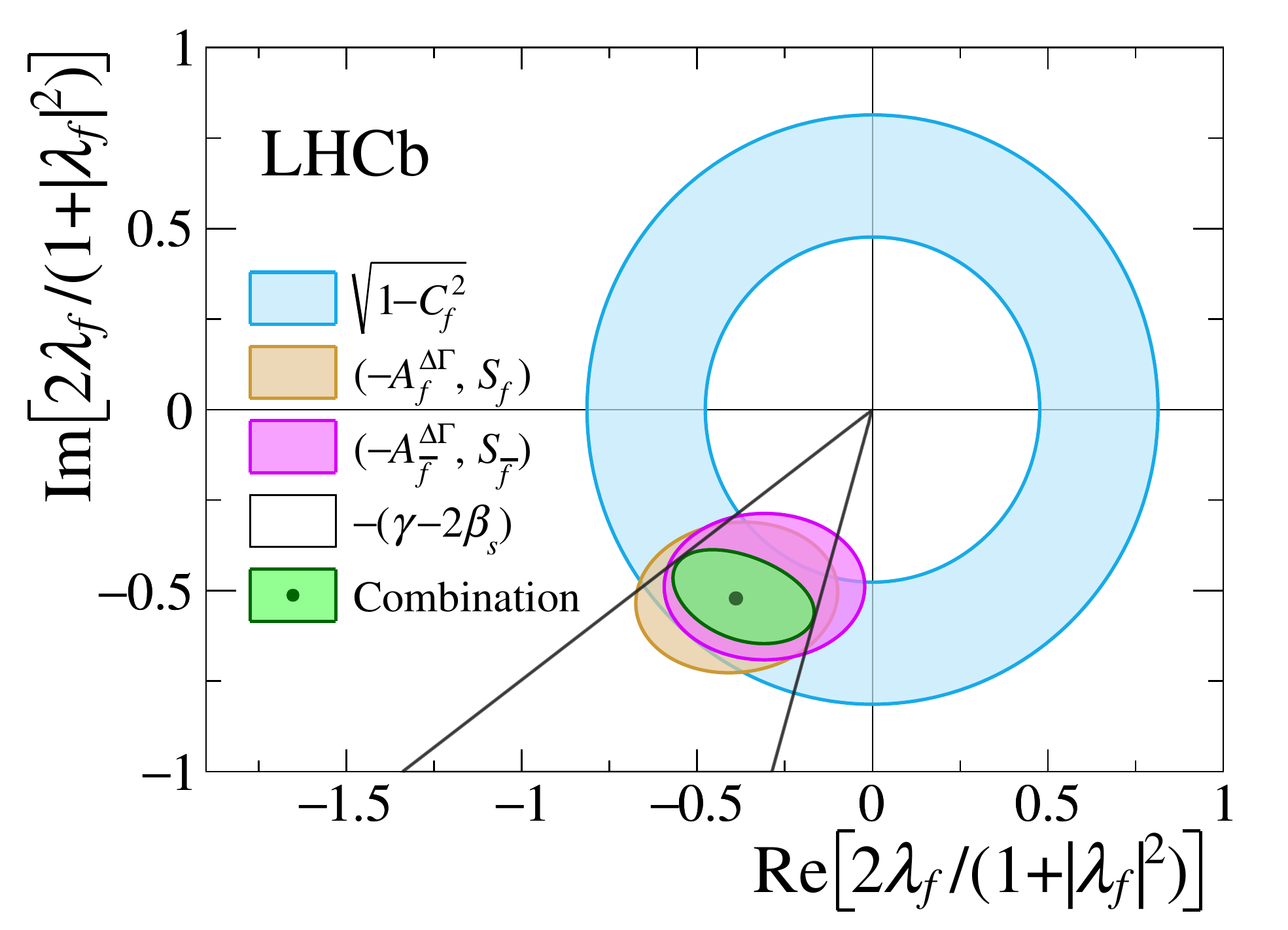}
\caption{
  \small
  Profile likelihood contours of (top left)
  \rdsk vs.~\g, and (top right)
  \strong vs.~\g.
  The markers denote the best-fit values.
  The contours correspond to 68.3\% CL (95.5\% CL).
  The graph on the bottom left shows \omcl for the angle \g, together with the central value
  and the 68.3\% CL interval as obtained from the frequentist method described
  in the text.
  The bottom right plot shows a visualisation of how each of the amplitude coefficients contributes towards the
  overall constraint on the weak phase, \weak.
  The difference between the phase of $(-\Dpar,\Spar)$ and $(-\Dbpar, \Sbpar)$ is proportional to the
  strong phase $\strong$, which is close to $360^\circ$ and thus not indicated in the figure.
}
\label{fig:interpretation_gamma}
\end{figure}

The value of $\beta_s$ is constrained to the value obtained from~\cite{HFAGSpring16},
$\phis = -0.030 \pm 0.033\rad$,
assuming $\phis = -2\beta_s$, \ie neglecting contributions from penguin-loop diagrams or
from processes beyond the SM.
The results are
\begin{align*}
\g      &= (128\,_{-22}^{+17})^\circ\,,\\
    \strong &= (  358\,_{-14}^{+13})^\circ\,,\\
\rdsk   &= 0.37\,_{-0.09}^{+0.10}\,,
\end{align*}
where the values for the angles are expressed modulo $180^\circ$.
Figure~\ref{fig:interpretation_gamma} shows the $1-\textrm{CL}$ curve for
\g, and the two-dimensional
contours of the profile likelihood $\mathcal L(\vec{\alpha}'_{\min})$.

The resulting value of \g is visualised in Fig.~\ref{fig:interpretation_gamma} by inspecting
the complex plane for the measured amplitude coefficients. The points determined by $(-\Dpar,\Spar)$ and $(-\Dbpar, \Sbpar)$ are
proportional to $\rdsk e^{i (\pm \strong - (\g - 2\beta_s))}$, whilst an additional constraint on \rdsk arises from \Cpar. 
The value of $\gamma$ measured in this analysis is compatible at the level of $2.3\,\sigma$, where $\sigma$ is the standard deviation, with
the value of $\gamma$ found from the combination of all LHCb measurements~\cite{LHCb-PAPER-2016-032} when all information from
\BsDsK decays is removed.
The observed change in the fit log-likelihood between the combined best fit point and the origin in the complex plane indicates $3.8\,\sigma$ evidence for \CP violation in $\Bs\to\Dsmp\Kpm$.

\section{Conclusion}
\label{sec:conclusion}
The \CP-violating parameters that describe the \BsDsK decay rates
have been measured using a data set corresponding to an integrated luminosity of $3.0\invfb$ of $pp$ collisions recorded with the \lhcb detector.
Their values are found to be
\begin{align*}
\Cpar  &= \phantom{-}0.73 \pm 0.14 \pm 0.05 \,, \\
\Dpar  &= \phantom{-}0.39 \pm 0.28 \pm 0.15\,, \\
\Dbpar &= \phantom{-}0.31 \pm 0.28 \pm 0.15\,, \\
\Spar  &=           -0.52 \pm 0.20 \pm 0.07\,, \\
\Sbpar &=           -0.49 \pm 0.20 \pm 0.07\,,
\end{align*}
where the first uncertainties are statistical and the second are systematic.
The results are used to determine the CKM angle \g, the strong-phase difference $\delta$ and the amplitude ratio $r_{\DsK}$ between the $B^{0}_{s}\to D_{s}^{-}K^{+}$ and $\Bsb \to D_{s}^{-}K^{+}$ amplitudes leading to
$\g = (128\,_{-22}^{+17})^\circ$, $\delta=(358\,^{+13}_{-14})^{\circ}$ and $r_{\DsK}=0.37\,^{+0.10}_{-0.09}$ (all angles are given modulo~$180^\circ$). This result
corresponds to $3.8\,\sigma$ evidence of \CP violation in this channel and represents the most precise determination of \g from \Bs meson decays.

\pagebreak

\addcontentsline{toc}{section}{References}
\setboolean{inbibliography}{true}
\bibliographystyle{LHCb}
\bibliography{main,LHCb-PAPER,LHCb-CONF,LHCb-DP,LHCb-TDR,ourbib}
\section*{Acknowledgements}
%
%
\noindent We express our gratitude to our colleagues in the CERN
accelerator departments for the excellent performance of the LHC. We
thank the technical and administrative staff at the LHCb
institutes. We acknowledge support from CERN and from the national
agencies: CAPES, CNPq, FAPERJ and FINEP (Brazil); MOST and NSFC
(China); CNRS/IN2P3 (France); BMBF, DFG and MPG (Germany); INFN
(Italy); NWO (The Netherlands); MNiSW and NCN (Poland); MEN/IFA
(Romania); MinES and FASO (Russia); MinECo (Spain); SNSF and SER
(Switzerland); NASU (Ukraine); STFC (United Kingdom); NSF (USA).  We
acknowledge the computing resources that are provided by CERN, IN2P3
(France), KIT and DESY (Germany), INFN (Italy), SURF (The
Netherlands), PIC (Spain), GridPP (United Kingdom), RRCKI and Yandex
LLC (Russia), CSCS (Switzerland), IFIN-HH (Romania), CBPF (Brazil),
PL-GRID (Poland) and OSC (USA). We are indebted to the communities
behind the multiple open-source software packages on which we depend.
Individual groups or members have received support from AvH Foundation
(Germany), EPLANET, Marie Sk\l{}odowska-Curie Actions and ERC
(European Union), ANR, Labex P2IO and OCEVU, and R\'{e}gion
Auvergne-Rh\^{o}ne-Alpes (France), RFBR, RSF and Yandex LLC (Russia),
GVA, XuntaGal and GENCAT (Spain), Herchel Smith Fund, the Royal
Society, the English-Speaking Union and the Leverhulme Trust (United
Kingdom).

\newpage
\centerline{\large\bf LHCb collaboration}
\begin{flushleft}
\small
R.~Aaij$^{40}$,
B.~Adeva$^{39}$,
M.~Adinolfi$^{48}$,
Z.~Ajaltouni$^{5}$,
S.~Akar$^{59}$,
J.~Albrecht$^{10}$,
F.~Alessio$^{40}$,
M.~Alexander$^{53}$,
A.~Alfonso~Albero$^{38}$,
S.~Ali$^{43}$
% ...

{\footnotesize \it
% ...
$ ^{5}$Clermont Universit{\'e}, Universit{\'e} Blaise Pascal, CNRS/IN2P3, LPC, Clermont-Ferrand, France\\
% ...
$ ^{10}$Fakult{\"a}t Physik, Technische Universit{\"a}t Dortmund, Dortmund, Germany\\
% ...
$ ^{38}$ICCUB, Universitat de Barcelona, Barcelona, Spain\\
$ ^{39}$Instituto Galego de F{\'\i}sica de Altas Enerx{\'\i}as (IGFAE), Universidade de Santiago de Compostela, Santiago de Compostela, Spain\\
$ ^{40}$European Organization for Nuclear Research (CERN), Geneva, Switzerland\\
% ...
$ ^{43}$Nikhef National Institute for Subatomic Physics, Amsterdam, The Netherlands\\
% ...
$ ^{48}$H.H. Wills Physics Laboratory, University of Bristol, Bristol, United Kingdom\\
% ...
$ ^{53}$School of Physics and Astronomy, University of Glasgow, Glasgow, United Kingdom\\
% ...
$ ^{59}$University of Cincinnati, Cincinnati, OH, United States\\
% ...
}
\end{flushleft}
\end{document}